%% file: template.tex
\documentclass{article}
\usepackage{arxiv}
\usepackage[utf8]{inputenc} 
\usepackage[T1]{fontenc}    
\usepackage{hyperref}       
\usepackage{url}            
\usepackage{booktabs}       
\usepackage{amsfonts}       
\usepackage{nicefrac}       
\usepackage{microtype}      
\usepackage{lipsum}		    
\usepackage{graphicx}
\usepackage{doi}
\usepackage{glossaries}
\usepackage[table]{xcolor}
\definecolor{green}{rgb}{0,.5,0}
\definecolor{magenta}{rgb}{.75,0,.75}
\usepackage{soul}
\sethlcolor{lightgray}
\usepackage[labelfont=bf,font=footnotesize]{caption}
\usepackage{float}
\usepackage{multicol}
\usepackage{wrapfig} 
\usepackage{csquotes} 
\usepackage{tikz}
\newcommand*{\rootpath}{}

\newcommand\scale{1} 
\newcommand\scaleppt{0.65}

\newcommand\corepluspenumbra{$\textrm{core}\cup\textrm{penumbra}$}
\newcommand\equationspacing{-5pt}
\newcommand{\review}[1]{\textcolor{black}{#1}}

\title{Final infarct prediction in acute ischemic stroke}
\renewcommand{\undertitle}{Introducing stroke, the core-penumbra concept and machine learning applications}
\renewcommand{\headeright}{}
\renewcommand{\shorttitle}{Final infarct prediction in acute ischemic stroke}
\author{
    \href{https://orcid.org/0000-0001-7206-2671}{\includegraphics[scale=0.06]{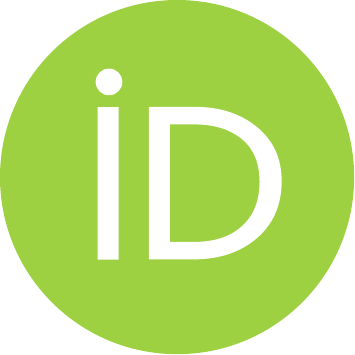}\hspace{1mm}Jeroen Bertels}\\
	Processing Speech and Images\\
	Department of Electrical Engineering\\
	KU Leuven, Belgium\\
	\texttt{jeroen.bertels@kuleuven.be}\\
	\And
	David Robben\\
	Processing Speech and Images\\
	Department of Electrical Engineering\\
	KU Leuven, Belgium\\
	\texttt{david.robben@kuleuven.be}\\
    \And
	Dirk Vandermeulen\\
	Processing Speech and Images\\
	Department of Electrical Engineering\\
	KU Leuven, Belgium\\
	\texttt{dirk.vandermeulen@kuleuven.be}\\
	\And
	Robin Lemmens\\
	Laboratory of Neurobiology\\
	Department of Neurosciences\\
	KU Leuven, Belgium\\
	\texttt{robin.lemmens@kuleuven.be}
}
\begin{document}
\date{}  
\maketitle
\begin{abstract}
	This article focuses on the control center of each human body: the brain. We will point out the pivotal role of the cerebral vasculature and how its complex mechanisms may vary between subjects. We then emphasize a specific acute pathological state, i.e., acute ischemic stroke, and show how medical imaging and its analysis can be used to define the treatment. We show how the core-penumbra concept is used in practice using mismatch criteria and how machine learning can be used to make predictions of the final infarct, either via deconvolution or convolutional neural networks.
\end{abstract}
\input{glossary.tex}
\section{The cerebral vasculature}
On average, the human brain accounts for about 20~\% of the total energy consumption and a similar fraction 
of the total cardiac output of an adult at rest, while the brain represents only about 2~\% of the whole body size~\cite{Mraovitch1996,Kety1950}. Unable to store energy and relying almost exclusively on the aerobic consumption of glucose, a constant functioning of the cerebral vasculature is paramount to carry oxygenated, glucose-rich blood throughout the brain parenchyma to every single cell.\\
From a hemodynamics standpoint, and under the assumption of having an adequate blood composition in the feeding arteries to the brain, parenchymal access to oxygen and glucose can be described by two time-dependent parameters: the cerebral blood flow (\gls{cbf}) and the cerebral blood volume (\gls{cbv}), i.e., the flow rate through (ml/100~g/min) and volume of blood in (ml/100~g) a certain piece of tissue, respectively. As such, the correct and continuous functioning of brain cells depends on the constant tuning of \gls{cbf} and \gls{cbv} at each location, keeping them within the functional range. Healthy values of \gls{cbf} and \gls{cbv} may shift slightly across subjects, and are different in gray matter (around 80~ml/100~g/min and 4~ml/100~g) and white matter (around 20~ml/100~g/min and 2~ml/100~g)~\cite{Shetty2006,Carroll2008}. The brain has developed anatomical and functional features to combat various pathological vascular conditions that would influence those variables~\cite{Markus2004}.
\subsection{Anatomy}
\input{\rootpath figures/vasculature}
The first line of defense lies in the anatomical structure of the cerebral vasculature (Figure~\ref{fig:vasculature}). Three large arteries feed oxygenated, glucose-rich blood to the brain, while two large veins drain deoxygenated, glucose-deprived blood from the brain. Inside the brain, these vessels form diverging and converging branches, respectively, and connect at the level of the capillaries, where eventually, the extracellular fluid mediates the exchange to and from the cells. Due to the well-developed collateral circulation, i.e., blood may follow different trajectories to reach a certain destination, \gls{cbf} and \gls{cbv} may be well preserved across the entire brain even in the case of an abrupt and local structural change. This way, the communication between the three feeding arteries at the circle of Willis is of crucial importance~\cite{Riggs1963}. Other significant collaterals exist, e.g., through anastomoses between the external carotid artery branches and the intracerebral circulation and between vessels at the surface of the brain~\cite{VanderEecken1953}. Nonetheless, some brain regions are particularly vulnerable, e.g., regions at the boundaries between the anterior, middle, and posterior cerebral arteries and territories supplied by end arteries~\cite{Markus2004}.
\subsection{Autoregulation}
As a second line of defense, nature has equipped the brain with the functional phenomenon of cerebral autoregulation. Under normal conditions, there is a direct relationship between \gls{cbf} and \gls{cbv}. The cardiac system maintains a negative pressure gradient, i.e. cerebral perfusion pressure (\gls{cpp}), across the brain from arterial input to venous output, enough to overcome the cerebrovascular resistance (\gls{cvr}) and reaching sufficient \gls{cbf} at each location (Figure~\ref{fig:vasculature}). When the small intracranial vessels (the predominant resistance vessels) constrict or dilate, the \gls{cvr} increases or decreases, respectively, and the \gls{cbv} and \gls{cbf} would both decrease or increase until new steady-state conditions are obtained. Via active or passive mechanisms of vasoconstriction and -dilation, \gls{cbf} remains relatively constant despite moderate variations in \gls{cpp}. 
\subsection{Other mechanisms}
To compensate for additional hemodynamic changes, brain parenchyma can also alter its own state and increase or decrease the oxygen extraction fraction (\gls{oef}) from blood~\cite{Markus2004} (Figure~\ref{fig:vasculature}). Numerous studies have shown that a local increase in neuronal activity results in a local increase in glucose utilization, accompanied by local increases in \gls{cbf}~\cite{Roy1890}. Changes in blood O$_2$ and CO$_2$ concentrations act through direct and indirect effects. In conditions of hypoxia or hypercapnia, via concomitant changes in the pH of brain tissue, \gls{cbf} is altered directly by changing the \gls{cvr} or indirectly through the release of vasoactive factors~\cite{Mraovitch1996}.
\subsection{Intra- and inter-subject variability}
The state and capacity of these defense mechanisms vary widely across subjects. Although we all share the basic cerebrovascular anatomy, collateral circulation can be more or less pronounced~\cite{Liebeskind2003}. Even under normal operating conditions, the lower and upper limits of cerebral autoregulation are not fixed. As such, the exact values of both \gls{cbf} and \gls{cbv} may vary due to various physiological variables, including arterial blood gases (e.g. NO is an important mediator controlling basal \gls{cbf}~\cite{White1998,White1999}) and metabolic rate through vasoneuronal coupling (e.g. an upward shift of both limits due to sympathic activation to anticipate an increase in \gls{cpp}~\cite{Markus2004}). Similarly, certain disease states result in a prolonged shift (e.g., an upward shift of the autoregulatory plateau in patients with chronic hypertension~\cite{Strandgaard1973}).
\section{Acute ischemic stroke}\label{sec:acute_ischemic_stroke}
Despite these mechanisms, there are pathological states of what is called cerebral \textit{ischemia} in which the \gls{cbf} and \gls{cbv} are insufficient to meet the metabolic demand. A chronic fall in \gls{cbf} can happen, e.g., under systemic arterial hypotension or severe stenosis in one of the feeding intra- or extracranial arteries. However, before the cerebral metabolic rate of oxygen begins to fall, the \gls{oef} increases as a fallback mechanism. Therefore, it is primarily an acute change in the cerebral vasculature that pushes the \gls{cbf} abruptly outside its operating range, albeit more likely in subjects with preliminary chronic changes and limited collateral circulation.
\subsection{Neuronal dysfunction}
In severe focal ischemia, the \gls{cbf} starts to decrease passively with any further decrease in \gls{cpp}. From that point we can identify critical \gls{cbf} thresholds at which certain neuronal functions are lost~\cite{Hossmann1994} (Figure~\ref{fig:thresholds}). This pattern of thresholds is complex, and the exact levels vary with the duration of the event and across different animal models. Generally, the first thresholds ($\pm$~80-40~ml/100~g/min) relate to a decline in protein synthesis, selective gene expression, and selective neuronal loss. A more progressive breach ($\pm$~40-15~ml/100~g/min) results in the loss of electrical function, transpiring into clinical symptoms. The sudden onset of symptoms of neurological deficit of presumed vascular origin results in the clinical diagnosis of stroke. Two main types of stroke can be defined based on the acute cause of the ischemia: an ischemic stroke ($\pm$~80~\%) and a hemorrhagic stroke ($\pm$~15~\%). In an acute ischemic stroke (\gls{ais}), the ischemia is attributed to the occlusion of an artery \review{(e.g, as a result of an embolism originating from the heart)}. In a hemorrhagic stroke, an artery is ruptured, resulting in the acute accumulation of blood within the brain parenchyma. Often, the classification includes a third type: a transient ischemic attack ($\pm$~5~\%). A transient ischemic attack is similar to an \gls{ais}, but as its name suggests, the ischemia is only temporary and does not cause permanent damage~\cite{Coutts2005}. The effects and potential treatments may differ between ischemic and hemorrhagic stroke. Nonetheless, they share a similar symptomatic pattern and progression of the parenchymal tissue under ischemia. \review{From now on, the focus is on AIS, more specifically following a large vessel occlusion in the anterior circulation.}
\input{\rootpath figures/thresholds}
\subsection{Core and penumbra}
When the \gls{cbf} decreases to the lowest regions ($\pm$~15-0~ml/100~g/min) a loss of cellular ion hemostasis occurs~\cite{Hossmann1994,VonKummer2017}. While the clinical symptoms are similar, brain tissue will typically undergo irreversible tissue damage and become \textit{infarcted}, i.e., the \textit{core} of the ischemic tissue. With the introduction of irreversible tissue damage, we have implicitly introduced reversible changes for higher levels of \gls{cbf}, thus brain tissue that may recover when the blood perfusion is restored, i.e., the \textit{penumbra} of the ischemic tissue. However, this whole concept of irreversible versus reversible, core versus penumbra, and non-salvageable versus salvageable, is highly complex and is still the subject of ongoing research. What is certain is that there is no one-to-one relationship with the absolute value of the \gls{cbf}, not across different subjects nor within a single subject. For an intuitive elaboration of what would happen when blood flow is restored, thus what part of the brain can be salvaged, we need to introduce at least two additional variables: time and location (Figure~\ref{fig:thresholds}).
\subsubsection{Time}
\input{\rootpath figures/progression}
Imagine that the immediate opening of the vessel follows the abrupt occlusion. In this case, tissue may recover completely, despite local, short-term levels of detrimental \gls{cbf}. Per definition, non-salvageable and salvageable tissue can only be defined at a certain moment after stroke onset (Figure~\ref{fig:progression}). In fact, the concept that neuronal survival depends on the time since onset (\gls{tso}) dates back from 1981~\cite{Astrup1981,Heiss1983,Jones1981}. From here a continuum emerged of neurons surviving indefinitely with a \gls{cbf} above 15~ml/100~g/min but dying within 30~min with a \gls{cbf} below 10~ml/100~g/min. In non-human primates, for instance, brain ischemia of 10 to 15~ml/100~g/min can sustain for 2 to 3 hours without irreversible injury. This ``time is brain'' relationship has a deep-rooted foundation in early clinical trials, which do patient triage based on the \gls{tso}~\cite{Marler1995,Goyal2016}. Since these treatments aim to restore blood perfusion, this implicitly states that the amount of salvageable tissue decreases over time and thus that the core eats away the penumbra.
\subsubsection{Location}
In a way, the fact that an increasing \gls{tso} results in a virtual, upward shift of the lowest \gls{cbf} region at a particular location can be attributed to the status of the neighboring parenchyma. This spatial dimension arises since a complex cascade of biochemical processes lies at the origin of cellular breakdown. With a progressive decrease in \gls{cbf} across the critical levels, the increase in anaerobic energy metabolism results in acidosis with an increase in lactate and glutamate. Eventually, neuronal cells will undergo a generalized collapse of membrane function with anoxic depolarisation due to unbalanced in- and efflux of cellular ions. It has been shown that this malign environment can spread out, e.g., via peri-infarct depolarizations~\cite{Hossmann1996}, diffusion of glutamate~\cite{Castillo1997} or inflammation-related processes~\cite{Vila2000}.
\subsection{Treatment}
At a certain point, the entire penumbra will turn into core (Figure~\ref{fig:progression}). As discussed before, the time this takes and the final volume vary widely across patients. Any intermediate external intervention that leads to increased blood perfusion can be considered a potential treatment. In practice, there are two established reperfusion therapies for an \gls{ais}: via the administration of drugs (i.e. intravenous treatment (\gls{ivt})~\cite{Marler1995}) or via \review{mechanical recanalization} (i.e. thrombectomy or intra-arterial treatment (\gls{iat})~\cite{Goyal2016}). Care should be taken to avoid distal microvascular obstruction, which would further contribute to the progression of the ischaemic damage~\cite{DelZoppo2000}. Different approaches to quantifying the effect of treatment are possible. Straightforwardly, the direct effect on blood perfusion can be quantified by the \review{modified} Thrombolysis in Cerebral Infarction (\gls{tici}) score~\cite{Higashida2003}. The higher the \gls{tici} score (\gls{tici} $\in [0,1,\textrm{2a},\textrm{2b},\textrm{2c},3]$), \review{the better the recanalization (potentially the reperfusion)} and the slower the further progression of the infarct (Figure~\ref{fig:progression}). Additionally, it is important for this effect to take place soon after hospitalization and when the decision on the treatment was made, i.e., the time to treatment (\gls{ttt}) (Figure~\ref{fig:progression}). Success also means avoiding adverse effects such as bleeding, cerebral swelling, or damage to the inner arterial lining (which happens during \gls{iat}). Other quantifiers correlate indirectly with the amount of reperfusion but aim to score a patient's well-being days or months after the event. As such, the National Institute of Health Stroke Scale score (\gls{nihss}) measures the neurological deficit by scoring the specific ability on 11 items. The higher the \gls{nihss} score (\gls{nihss} $\in [0,42]$), the larger the impairment. The modified Rankin Scale (\gls{mrs}) score measures the degree of disability or dependence in daily activities. The larger the \gls{mrs} score (\gls{mrs} $\in [0,6]$), the larger the disability, with 0 having no symptoms and 6 indicating dead. The \gls{mrs} has become the most widely used clinical outcome measure for evaluating potential stroke therapies in clinical trials.
\section{The role of imaging}\label{sec:imaging}
The likelihood of recovery to independence after \gls{ais} is significantly improved by reperfusion, either by \gls{ivt} or \gls{iat}. However, the marginal correlation between \gls{tici} and \gls{mrs} is weak due to the highly complex interaction between subject, time, location, and treatment. There are certain scenarios in which treatment should be avoided. In general, this treatment decision depends on balancing the possibility of a good clinical outcome against the risk of complications. Complications of \gls{ivt} include intracranial hemorrhage (\gls{ich}), systemic bleeding and orolingual oedema~\cite{Fugate2015,Tawil2017}. Complications of \gls{iat} also include \gls{ich}, but are mostly related to radiological contrast media and device-related vascular injury~\cite{Tawil2017}. The risk of \gls{ich} is directly related to the ischemia status since oxygen and glucose are also needed to maintain the structural integrity of the vessels. Therefore, the absolute estimation of core and penumbra volumes at the time of triage and the time of estimated reperfusion becomes the key ingredient for making a well-considered decision (Figure~\ref{fig:progression}). It is for this purpose that acute imaging is used.
\subsection{Acute imaging}
Two complementary imaging approaches can be classified based on the direct or indirect nature of estimating the underlying parenchymal tissue status. Direct methods rely on the fact that the underlying biochemical cascade changes the measured signal of a particular imaging modality at the voxel level. Indirect methods do this via the offline measurement of blood perfusion dynamics from a series of direct image measures at the voxel level. Due to partial volume effects, the measured voxel value represents a sort of average of the underlying mechanisms and interactions that are taking place between the tissue and the imaging signal, which is then captured by the imaging detector and finally reconstructed as a voxelized image. In practice, \review{the voxel size could be} 1~mm~x~1~mm~x~1~mm, defined as a compromise between image noise, resolution, and acquisition time. If we consider the cuboid shown in Figure~\ref{fig:vasculature} to represent a single voxel of that size, it could be a network of hundreds of capillaries and thousands of cells, potentially from multiple capillary beds, fed and drained by different arterioles and venules, respectively.
\subsubsection{Direct imaging}
Via the extensive histological study of the evolution of ischemia in animal models, it became clear that early molecular changes, corresponding to the higher \gls{cbf} threshold levels and even selective neuronal necrosis, are unlikely to be detected by practical imaging methods~\cite{Garcia1983,Garcia1984,Garcia1988,Garcia1993,Garcia1995,Garcia1997}. A better imaging biomarker is ischemic edema, i.e., monitoring the intra- and extra-cellular water content. It turns out that the complex cascade of cellular breakdown, as soon as there is energy depletion, goes hand in hand with different types of ischemic edema~\cite{Simard2007,VonKummer2017}.  First, cytotoxic or cellular edema results in the redistribution of ions and water from the extra- to the intra-cellular environment (starting from \gls{cbf} values below $\pm$~30~ml/100~g/min~\cite{Hossmann2006}). Then, when there is (minimal) perfusion and an intact blood-brain-barrier, ionic edema results in both the redistribution of ions and water from the intra-vascular to the extra-cellular environment and the net increase of tissue water content (starting from \gls{cbf} values below $\pm$~20~ml/100~g/min~\cite{Symon1979}), compensating for changes induced by the cytotoxic edema. The ionic edema itself may further decrease the \gls{cbf}, and depending on the exact level and duration of the event, lead to irreversible changes within a timespan of 60~min~\cite{Ito1979,Todd1986}. Finally, vasogenic edema results in a further net increase of tissue water content due to blood-brain-barrier failure, followed by leakage of plasma proteins. This will result in hydrostatic and osmotic pressure gradients, causing severe brain tissue swelling, mass effects, and a further reduction in \gls{cpp}, resulting in irreversible damage. Yet again, care must be taken with timing and absolute \gls{cbf} levels due to the complex interplay between time and location.
\paragraph{\gls{ncct}}
In one voxel, the overall attenuation coefficient for X-rays will only change as soon as there is a net increase in tissue water content. As a result, it is evident, and confirmed with experimental and clinical observations~\cite{VonKummer2017}, that plain computed tomography (\gls{ct}), i.e. non-enhanced or non-contrast \gls{ct} (\gls{ncct}), can detect ionic brain edema and as such irreversible ischemia. However, early ionic changes, with an increase in tissue water content of less than 1~\%, result in subtle hypoattenuation, especially in gray matter. Expert knowledge is required, and considerable inter-rater variability remains in voxel-based analysis. In an attempt to quantify the \gls{ct} hypoattenuation, and thus ionic edema, within the \gls{mca} territory, the Alberta Stroke Program Early \gls{ct} Score (\gls{aspects}) was introduced~\cite{Barber2000,Puetz2009}. The lower the \gls{aspects} (\gls{aspects} $\in [0,10]$), the more regions are included in the segmental estimate of the infarct, with 10 indicating no regions and 0 indicating all 10 predefined regions involved.
\paragraph{\gls{mri}}
Frequently used \gls{mri} sequences for \gls{ais} imaging fall under the term of diffusion-weighted imaging (\gls{dwi}), in which the diffusion of water molecules interacts with the imaging signal and creates contrast. 
Since cytotoxic edema is associated with diffusion impairment, it will lead to \review{hyperintensity} on \gls{dwi}. For quantification purposes, the apparent diffusion coefficient (\gls{adc}) in each voxel, mainly influenced by Brownian motion, can be calculated from \gls{dwi} values at different settings. Indeed, it has been shown that \gls{adc} values decrease as soon as \gls{cbf} drops below 20-40~ml/100~g/min, on its turn related to shrinkage of extracellular space~\cite{Lin2003,Wang2000,Hossmann2006}.
\subsubsection{Indirect imaging}\label{sec:indirect_imaging}
As discussed, blood perfusion dynamics, such as \gls{cbf}, greatly influences the tissue status. Indirect methods, therefore, aim to capture these dynamics on the voxel level and relate this to reversible or irreversible damage.
\paragraph{\gls{ctp} and \gls{pwi}}
\input{\rootpath figures/deconvolution}
The idea is simple: after the intravenous injection of a contrast agent at t\textsubscript{0} a series of \gls{3d} images, acquired between t\textsubscript{s} and t\textsubscript{e}, show the passage of contrast through the brain (Figure~\ref{fig:deconvolution}). For this purpose, either \gls{ct} perfusion (\gls{ctp}) or perfusion-weighted imaging (\gls{pwi}) can be used, respectively using \gls{ct} and \gls{mri} imaging to capture the change of contrast inside a piece of tissue (remember the cuboid from Figure~\ref{fig:vasculature}) as a change in image intensity, on its turn converted into a change in concentration of the contrast agent over time, i.e. the time concentration curve (\gls{tcc}). Nevertheless, even for a piece of ``tissue'' fully inside a vessel, the \gls{tcc} will flatten along its trajectory in the vessel due to laminar flow and diffusion. Depending on the exact situation (e.g., the injection protocol itself, a patient's cardiovascular system, motion), the \gls{tcc}s will look different across different patients and multiple measurements. In theory, it is possible to normalize cerebral \gls{tcc} measurements as if they were all the result of an identical \gls{tcc} at the input of the brain, i.e., the arterial input function (\gls{aif}). Deconvolving the measured \gls{tcc}s with the measured \gls{aif} normalizes the \gls{aif} to be the Dirac impulse (\gls{aif}\textsubscript{\gls{imp}}) and normalizes the \gls{tcc}s to be the tissue's impulse response functions (\gls{tcc}\textsubscript{\gls{irf}}). It turns out that, under reasonable assumptions (e.g., the contrast agent stays within the vessels), this deconvolution strategy is indeed valid and that from the \gls{tcc}\textsubscript{\gls{irf}} certain perfusion dynamics can be derived straightforwardly.
\paragraph{\gls{pet} and \gls{spect}}
In contrast to the endogenous character of \gls{ctp} and \gls{pwi}, nuclear medicine methods use exogenous radionuclides as tracers. In both single photon emission CT \gls{spect} and positron emission tomography \gls{pet} the accumulation and decumulation of these tracers inside the tissue can be measured, from which the blood perfusion dynamics can be derived~\cite{Yonas1996}. The fully quantitative nature of \gls{pet} further allows the measurement of the \gls{oef}. Currently, \gls{pet} and \gls{spect} are used less frequently compared to \gls{ctp} and \gls{pwi} in the acute clinical setting. Nonetheless, they deserve mentioning due to their early contribution to pathophysiological research on irreversible versus reversible tissue damage following an \gls{ais}~\cite{Baron1981,Baron1999,Wise1983,Marchal1996}.
\subsubsection{Other imaging}
Besides imaging that tries to estimate tissue status on a voxel basis, additional imaging is needed to triage patients correctly because \gls{iat} requires \review{mechanical recanalization}, which limits its application to proximal thrombi.
\paragraph{\gls{cta} and \gls{mra}}
\gls{ct} angiography (\gls{cta}) or \gls{mri} angiography (\gls{mra}) aims to visualize the vasculature in the entire head and neck region with sufficient resolution, primarily to assess the exact location of the occlusion and its mechanical accessibility to effectively perform \gls{iat}. The acquisition is similar to \gls{ctp} and \gls{pwi}, with the machine parameters and contrast injection tuned to obtain a superior axial extent and image quality at the expense of acquiring only a single \gls{3d} volume. Despite its primary objective, \gls{cta} or \gls{mra} is also used to assess collateral circulation (\gls{cc}) scores in a qualitative manner. For example, the degree of pial or leptomeningeal \gls{cc} in \gls{ais} has been associated with infarct volume, \gls{ich} and \gls{mrs}~\cite{Liebeskind2003,Christoforidis2009,Bang2011,Maas2009,Tan2009}. More recently, multi-phase \gls{cta} or \gls{mra} is used, e.g., to visualize the vasculature during early-arterial, late-arterial, and venous passage of contrast agent. In this perspective, multi-phase \gls{cta} may improve \gls{cc} evaluation, showing superior inter-rater agreement and clinical outcome prediction~\cite{Menon2013a,Menon2015a,Nambiar2014}. More recently, it has been shown that \gls{iat} is effective after \gls{cc} score-based patient selection~\cite{Goyal2015} and that \gls{cc} scores can independently predict the progression of middle cerebral artery infarct~\cite{Flores2015}.
\paragraph{\gls{ctp} versus \gls{cta}}
Technical limitations in modern \gls{ct}, such as the rotation speed ($\pm~360^{\circ}$/s) and axial detector size (4-16~cm), result in a trade-off between the temporal resolution and the axial extent. Detector movements in the axial direction, via helical or jog-mode scanning protocols, increase the axial extent at the cost of temporal resolution and additional processing (e.g., motion correction, different axial positions have different timings). Furthermore, multiple acquisitions at a certain axial position increase radiation dose. Following the ``as low as reasonably possible'' principle, this further adds the tube-load and the total number of acquisitions to the trade-off. The combination of (i) a practical \gls{ctp} iodine contrast injection protocol (5-6~ml/s for a duration of 5-6~s), (ii) variations in the cardiovascular dynamics taking place between the injection site and the brain inlet, and (iii) variations in cerebral perfusion dynamics, demand a \gls{ctp} duration t\textsubscript{e}-t\textsubscript{s} of about 1~min (Figure~\ref{fig:deconvolution}). Recent research also investigated the effects of varying the temporal resolution and tube load on the diagnostic quality of \gls{ctp}-derived perfusion dynamics~\cite{Ioannidis2021}. The lowest possible dose protocol was set at $\Delta$t~=~2~s and a tube load of 100~mAs, which limits the axial extent of current \gls{ctp} to 8-32~cm as observed in clinical practice. Given these variables, \gls{cta} is nothing more than a specific \gls{ctp} setting with (i) a wider injection bolus ($\pm$~3~ml/s for a duration of $\pm$~25~s), (ii) $\Delta$t~=~t\textsubscript{e}-t\textsubscript{s} and (iii) t\textsubscript{e}-t\textsubscript{s} such that the entire head and neck region is captured (Figure~\ref{fig:deconvolution}). Since the reconstruction of a single rotation averages the underlying dynamics of the contrast agent over the rotation period, the exact setting is defined such that the maximum of any vascular \gls{tcc} plateaus over the entire scan duration. It has been suggested that multi-phase \gls{cta} could substitute the diagnostic value of \gls{ctp}. Though this might be true from a pure diagnostic standpoint, in the limit \gls{ctp} contains (multi-phase) \gls{cta}, while exact blood perfusion dynamics can only be obtained using higher temporal sampling rates.
\subsection{Follow-up imaging}
Follow-up imaging aims to reassess the state of the cerebral tissue and check if the treatment was successful. Especially in the case of \gls{ivt}, guidelines recommend a follow-up \gls{ct} or \gls{mri} scan after 24~h to exclude \gls{ich} and decide on the initiation of a secondary preventive treatment with antithrombotic agents~\cite{Powers2018}. In general, without any secondary reperfusion treatment, follow-up imaging should be able to indicate the final infarct. As a result, this type of follow-up imaging should be acquired no sooner than when the infarct has fully progressed (Figure~\ref{fig:progression}). Typical timing is between 24~h and 5~days. The type of imaging can be similar to the acute phase, mostly direct since the more pronounced edema can now be detected with higher sensitivities.
\section{Final infarct prediction}\label{sec:introduction}
Estimates of core and penumbra volumes can be seen as imaging biomarkers. Integrated with clinical information of the patient (e.g. age, stroke history, drugs), an estimated \gls{ttt} and \gls{tici} score, they form a holistic basis to triage \gls{ais} patients. For this purpose, recent research and trials focus on the quantitative and voxel-based estimation of core and penumbra using data-driven methods.
\subsection{Data-driven core and penumbra estimation}\label{sec:datadriven}
Data-driven methods use machine learning to learn a mapping from input to output (Figure~\ref{fig:machine_learning}). This mapping is accomplished by a certain model, which contains a number of parameters that are tuned such that the mapping is optimal on a known dataset of input-output pairs. During optimization, the difference between a ground truth (e.g. manual) output and its prediction, i.e. the ``loss'', is minimized by iterating across all input-output pairs in the dataset until convergence. State-of-the-art methods aim to produce segmentation maps, e.g. one for the core and one for the penumbra. Before digging into the current benchmark, we will describe briefly two fundamental aspects regarding the core and penumbra terminology.
\input{\rootpath figures/machine_learning}
\subsubsection{Ground truth generation}
While the idea to learn a data-driven mapping is simple, it \review{raises} the question of how to obtain valid datasets of input-output pairs for the tasks of core and penumbra estimation. Following Section~\ref{sec:imaging}, it is expected that the core and the penumbra can be deduced from acute \gls{ncct} and \gls{dwi} imaging, respectively. However, there are two limitations to their effective use. First, in clinical practice the concurrent acquisition of both modalities is rare, and most often a sole \gls{ct}-based workup is done due to various reasons, including availability, safety, and speed. Second, the exact mapping from \gls{ct} or \gls{adc} values to irreversible or reversible changes, respectively, is non-trivial. Therefore, from a theoretical point of view, follow-up imaging, either \gls{ct}- or \gls{mri}-based, is the gold standard. In fact, by using follow-up imaging we construct a database in which the output represents the follow-up, i.e. final, infarcts. In the limits, in some subjects the final core will be equal to the acute core, i.e. the core at the time of acute imaging, in others it will represent the acute core plus penumbra (\corepluspenumbra). Respectively, these groups are often called \review{reperfusers and non-reperfusers}, thus reflecting subjects with instantaneous complete reperfusion at the time of acute imaging and subjects in whom no (partial) reperfusion was achieved until follow-up.
\subsubsection{Ambiguities in their definition}\label{sec:whatiscore}
Up till now, we followed the definition of core and penumbra as being, at a certain point in time, irreversibly and reversibly damaged, and thus non-salvageable and salvageable tissue, respectively. From a practical point of view, it is impossible to construct a database as described before for acute core estimation since it is impossible to obtain complete recanalization at the exact time of the acute imaging, i.e. there will always be a certain \gls{ttt}~>~0. Similarly, the description of the penumbra becomes ambiguous, while the \corepluspenumbra\ remains exact. This discrepancy has led to a secondary interpretation of their definition: core and penumbra as being the estimated irreversibly and reversibly damaged tissue at the time of recanalization~\cite{Goyal2020}. In other words, a medical doctor often reasons in terms of ``What can \textit{we} still save?", inherently taking into account a significant \gls{ttt}. Potentially, this has led to a misinterpretation of the value of \gls{dwi}, which is generally accepted as the gold standard to identify the core, and thus irreversibly damaged tissue. However, this is only true as far as it concerns the estimation under current constraints, and as such, care must be taken when interpreting the acute \gls{dwi}. For example, if \gls{dwi} is (partly) sensitive to the penumbra, it could be the modality of choice to estimate follow-up infarct volumes when \gls{ttt}~>~0.
\subsubsection{The present: \gls{dcv}-based estimation}\label{sec:features}
Most data-driven methods to estimate core and penumbra in clinical practice use indirect imaging, more specifically perfusion imaging. In fact, via \gls{dcv}, perfusion measurements can be normalized and quantified across different subjects and acquisitions, and the resulting \gls{tcc}\textsubscript{\gls{irf}}s contain direct information on the blood perfusion dynamics of the underlying parenchymal tissue~\cite{Fieselmann2011}. For example \gls{tmax} and \gls{cbf} maps can be derived from the \gls{tcc}\textsubscript{\gls{irf}} map as:
\begin{align}
    \textrm{\gls{tmax}} &= \arg\max_{t}\textrm{\gls{tcc}}_{\textrm{\gls{irf}}},\\[\equationspacing]
    \textrm{\gls{cbf}} &\propto \max_{t}\textrm{\gls{tcc}}_{\textrm{\gls{irf}}}.
\end{align}
After dichotomization into input-output datasets of \review{reperfusers and non-reperfusers}, the typical strategy is to find the optimal combination of perfusion parameter and threshold with respect to volume estimation of the follow-up infarct~\cite{Wintermark2006,Bivard2013}. It turns out that the follow-up infarct volume of \review{reperfusers and non-reperfusers}, and thus the acute core and \corepluspenumbra\ is best approximated with:
\begin{align}
    \textrm{\corepluspenumbra} \approx \textrm{\gls{tmax}} &\geq 6~\textrm{s};\\[\equationspacing]
    \textrm{core} \approx \textrm{\gls{rcbf}} &< 0.30,
\end{align}
as in~\cite{Olivot2009} and~\cite{Campbell2011}, respectively. Note that instead of the \gls{cbf} the relative \gls{cbf} (\gls{rcbf}) is used, in which the \gls{cbf} values are relative to normal white matter. The use of \gls{tmax} and \gls{rcbf} to obtain core and penumbra estimates can be considered the current benchmark. Nevertheless, alternative perfusion maps and thresholds exist, e.g. using the relative \gls{cbv} (\gls{rcbv}) and delay time \review{(i.e., the \gls{tmax} after delay-corrected perfusion analysis)}, respectively \gls{rcbv} < 0.60~\cite{Cereda2016} and a delay time <~3~s~\cite{Lin2016} for core and \corepluspenumbra. While the core is often estimated using \gls{dwi}, the \corepluspenumbra\ is mostly estimated using perfusion imaging, hence the name perfusion lesion.
\subsection{Triage in clinical practice}
The importance of the imaging-based selection of patients that would benefit from acute reperfusion therapy grows rapidly~\cite{Scheldeman2021}. Nevertheless, their effective use to triage \gls{ais} patients is fairly new. In fact, clinical practices often still use qualitative proxies of tissue status derived from \gls{ncct} or \gls{mri}~\cite{Pexman2001} in combination with \gls{tso} windows. Luckily, the ongoing quest to select patients more carefully, both inside and outside these \gls{tso} windows, has led to the incorporation of so-called mismatch criteria into current clinical practice.
\subsubsection{Qualitative proxies and \gls{tso}}
Back in 1995, \gls{ivt} became the first internationally approved treatment for eligible \gls{ais} patients, at least when it could be administered within a \gls{tso} < 3~h~\cite{Marler1995}. It is only in 2008 that its efficacy and safety could be established in the extended time window of \gls{tso} < 4.5~h, in part due to the first (qualitative) imaging-based exclusion criteria, being a visible indication that > 1/3 of the \gls{mca} was affected~\cite{Hacke2008}. Similarly, in the triage for \gls{ivt} an \gls{aspects} < 7 was often used as contra-indication. This roughly relates to core estimation: the more irreversibly damaged tissue, the higher the risk for developing symptomatic \gls{ich} when blood perfusion would be restored. In addition, clinical deficit, e.g. in terms of the \gls{nihss} score, could help to make a rough estimate of the \corepluspenumbra, and thus the potential benefits.\\
The clinical introduction of \gls{iat} dates back from 2015 when five trials (MR CLEAN~\cite{Berkhemer2015}, ESCAPE~\cite{Goyal2015}, REVASCAT~\cite{Saver2015}, SWIFT PRIME~\cite{Campbell2015} and EXTEND IA~\cite{Jovin2015}) showed a strong advantageous treatment effect within a time window of \gls{tso} < 6~h in the HERMES meta analysis~\cite{Goyal2016}. The triage for \gls{iat} became independent to the triage for \gls{ivt}, and also incorporated rather simple and qualitative imaging-based exclusion criteria. In the triage of \gls{iat}, the exact role of these scores was not consistent. Some trials simply used an \gls{aspects} < 6 as contra-indication, mainly to reduce the risk for symptomatic \gls{ich}. Other trials used a multi-phase \gls{cta}-based \gls{cc} score~\cite{Goyal2015}.\\
While this qualitative triaging system is simple, it remains questionable since (i) reversible and early changes are not visible on \gls{ncct}~\cite{VonKummer2017}, (ii) those core and penumbra estimations are nor quantitative, nor voxel-based, and thus inefficient for long-term or location-based analyses, and (iii) it is prone to significant inter-rater variability~\cite{Farzin2016,McTaggart2015}. In this respect, the automated scoring of \gls{aspects} on acute \gls{ncct} has been investigated~\cite{Kuang2019,Hampton-till2015}, and proved to be non-inferior to their manual counterparts when correlated with \gls{dwi}-based follow-up infarct volumes~\cite{Nagel2017,Herweh2016}. By design, it cancels inter-rater variability, however, it still does not provide information on the tissue status of each voxel.
\subsubsection{Mismatch criteria}
Unfortunately, even though \gls{ivt} and \gls{iat} can be considered highly effective, the effect is mitigated by the low proportion of patients that eventually qualify for such treatments~\cite{AguiardeSousa2019}. For a true patient-specific triage, \gls{tso} should be seen as only one of many factors playing a role in lesion growth~\cite{Wouters2016,Menon2013,Yaghi2021,Ribo2007}. For this purpose, present research extends the use of qualitative \gls{ct}-based analysis with multi-modal biomarkers in the form of so-called mismatch criteria, i.e. how much does a lesion identified on one modality differ from a lesion identified on another modality (or even within the same modality), hence the name mismatch. A mismatch criterion is typically defined by a combination of the volumetric difference, the mismatch ratio and a constraint on one of the lesion volumes. It should be clear that an estimated core-penumbra mismatch can be obtained using different modalities, but all have the purpose to balance the risks and benefits of the treatment.\\
The \gls{dwi}-\gls{pwi} mismatch (i.e. difference > 10~ml, ratio > 1.2) was one of the first quantitative estimators to identify subgroups that were likely to benefit or to be harmed when administering \gls{ivt} in the extended \gls{tso} window of 3 < \gls{tso} < 6~h~\cite{Albers2006}. To a certain extent, the \gls{dwi}-\gls{pwi} mismatch was able to predict malignant mismatches, e.g. that would result in symptomatic \gls{ich}. Some criteria were used merely as a simple proxy for \gls{tso} when unknown (e.g. \gls{dwi}-\gls{flair}~\cite{Thomalla2011}). More recently, either direct or meta-analyses from WAKE-UP~\cite{Thomalla2018}, EXTEND~\cite{Ma2019} and ECASS-4: EXTEND~\cite{Campbell2019} have shown positive treatment effects of \gls{ivt} in the later or unknown \gls{tso} windows when either \gls{rcbf}-\gls{tmax} (i.e. mismatch > 10-20~ml, ratio > 1.2, core < 70-100~ml) or \gls{dwi}-\gls{flair} mismatch (visible mismatch, no \gls{flair} lesion) criteria were used~\cite{Thomalla2020}. Only few trials, e.g. THAWS~\cite{Koga2020}, were unable to show this effect\review{, potentially due to the use of lower-dose-\gls{ivt}}.\\
Likewise for \gls{iat}, DAWN~\cite{Nogueira2017} and DEFUSE 3~\cite{Albers2018} explored the use of core-penumbra mismatch, respectively \review{\gls{rcbf}/\gls{dwi}-\gls{nihss}} and \gls{dwi}-\gls{pwi}, in the unknown and later time window of \gls{tso} < 24~h. Both mismatches were able to show a positive treatment effect (mismatch > 15~ml, ratio > 1.8, core < 70~ml).
\subsection{The (near) future: CNN-based estimation}
It is clear that a quantitative and voxel-based estimation of tissue status can lead to superior patient triage, thereby avoiding the sole use of proxies such as \gls{tso} (if known) and other qualitative estimators as sole predictors of \gls{ais} progression. By doing so, the difference in \gls{ais} progression across patients would become part of the solution instead of presenting itself as an additional source of variation when analyzing treatment effects. Nevertheless, the \gls{dcv}-based estimation benchmark in current clinical practice has some limitations. Also, in parallel, more recent data-driven methods have been developed that are potentially more suitable for the task.
\subsubsection{Current limitations}
The problem with \gls{dcv}-based core and penumbra estimation is at least twofold. First, the method is generally considered significantly sensitive to noise~\cite{Boutelier2012,Meijs2016}, which is intrinsically relevant in \gls{ctp}. Both the relatively low temporal resolution and signal-to-noise ratio make the discrete \gls{dcv} mathematically ill-posed and hungry for regularization. In practice, both regularization and noise reduction are needed in order to obtain physiologically meaningful results~\cite{Fieselmann2011}. Furthermore, \gls{dcv} depends on \gls{aif} selection, which is prone to inter-rater variability~\cite{Robben2020}. Research towards automated \gls{aif} detection often targets local \gls{aif}s, which would be better suited for the theoretical \gls{dcv} model~\cite{Lorenz2006}. However, this further adds partial volume effects to the equation, which in turn thrives research towards (automated) venous output function integration with problems of its own~\cite{Lee2008a}.\\ 
Second, from Section~\ref{sec:acute_ischemic_stroke} it was clear that voxel-wise perfusion parameters alone per definition do not contain the necessary information to derive tissue status. As a result, the optimal solution for a single-parameter threshold model is very much dataset-dependent. It is not surprising that the optimal thresholds for derived perfusion parameters show large variability across different research~\cite{Bivard2013,Cereda2016}. Add to this the specific regularization, noise reduction and other preprocessing steps that typically vary across different vendors. So even when a fixed threshold is used for mismatch calculation, be careful, e.g. in~\cite{Fahmi2012} two software packages give differences of 23.6~ml and 15.8~ml for core and penumbra estimation, respectively.
\subsubsection{Added value of \gls{cnn}s}
Given these limitations, a first attempt using a generalized linear model to replace the single-parameter thresholding and including both \gls{pwi} and \gls{dwi} imaging had only limited success~\cite{Wu2001}. Later, non-linear models and region-based features were added to~\cite{Scalzo2012}. Similarly, after dichotomization into \review{reperfusers and non-reperfusers}, the use of random forest classifiers in combination with a larger spatial context and multi-modal \gls{mri} provided a substantial improvement over predefined thresholds to predict the follow-up infarct, respectively core and penumbra~\cite{McKinley2017}. In parallel, also research towards \gls{dcv}-free parameter estimation remained popular~\cite{Meijs2016}, but yet again seemed to ignore the effect of infarct location on its surroundings.
\paragraph{Spatial context and model complexity}
The abrupt takeover of almost the entire computer vision field by \gls{cnn}s, is no less true here. Following the ISLES 2017 challenge~\cite{Winzeck2018}, which compared methods to predict the follow-up infarct based on acute \gls{dwi} and \gls{pwi}, it was clear that \gls{cnn}-based methods were simply performing superior. Despite most methods still relying on perfusion parameters at the input, the model now incorporated spatiality and complexity more naturally. For example, deep \gls{cnn}s outperformed both shallower \gls{cnn}s and generalized linear models using nine biomarkers derived from \gls{mri}~\cite{Nielsen2018}. To test if the model could learn the treatment effect of \gls{ivt} they dichotomized the dataset correspondingly and two independent models were learnt. While most methods work purely voxel-based, some do implement additional shape space interpolation~\cite{Lucas2018}.\\
Most methods use perfusion maps as input or even include \gls{cnn}s in an end-to-end \gls{dcv}-based framework to generate the optimal \gls{aif}~\cite{DelaRosa2020}. Related to this, there are methods that use \gls{cnn}s with the sole purpose of directly estimating the perfusion maps, as such being able to overcome only some \review{practical} limitations of a \gls{dcv}-based framework~\cite{Ho2016,Hess2018,Robben2018a}. The use of native \gls{pwi} or \gls{ctp} as input is fairly new. Some methods use it as the sole input~\cite{Robben2020,Bertels2019a}, while most methods use it still in combination with other imaging such as \gls{dwi} or the derived perfusion maps~\cite{Pinto2018a}. The latter approach, in combination with adversarial training, resulted in top-performing methods in the ISLES 2018 challenge~\cite{Liu2019b,Chen2019}.
\paragraph{Incorporating metadata}
Putting pieces together, we know that apart from unleashing spatial context and model complexity on the imaging data, there is still non-imaging data, i.e. metadata, waiting impatiently to be incorporated. The dichotomization into \review{reperfusers and non-reperfusers} can be considered a naive way to incorporate \review{the effects of the \gls{tici} score and \gls{o24h}}, thereby producing two independent models that can deliver two virtual predictions, one for the core and one for the penumbra, respectively. This idea was similar to dichotomization based on \gls{ivt} administration~\cite{Nielsen2018,Wu2006}. However, for example, the \gls{ttt} of each subject within each group differs, and this while the growth rate of the core has been shown to vary widely across different subjects~\cite{Wheeler2015a,Guenego2018}. Similarly, recent \gls{dcv}-based research has shown that there is a clear positive correlation between \gls{ttt} and optimal \gls{rcbf} threshold for core volume estimation~\cite{DEsterre2015}. Thus in general it would be interesting to incorporate more metadata in the model more naturally. In this respect, a first attempt was made to train a multi-variate generalized linear model that used both the \gls{ctp} and clinical data to quantify the patient-specific dynamic change of tissue infarction depending on the \gls{ttt} and \gls{tici} score~\cite{Kemmling2015}. In a \gls{cnn}-based setup, \gls{tso}, \gls{ttt}, \gls{tici} score and final recanalization status (\gls{o24h}, i.e. is there an occlusion visible on $\pm$~24~h follow-up imaging) were integrated beautifully and shown to have the desired effect~\cite{Robben2020,Wouters2021}.
\section{Conclusion}
The brain hosts an intricate vasculature that is equipped with a variety of mechanisms to prevent ischemia. However, following an \gls{ais} the cerebral perfusion changes such that tissue becomes irreversibly or reversibly damaged, the core and the penumbra, respectively. Due to variability in defense mechanisms across different subjects and the complex interplay between time and location, the progression of the infarct is non-trivial. Either direct or indirect imaging methods can be used for patient triage to estimate the amount of salvageable tissue and assess the risk for potential complications. In clinical practice, the \gls{dcv}-based triage via core-penumbra estimation still dominates. More specifically, thresholding of \gls{ctp}-derived \gls{rcbf} and \gls{tmax} maps for core and \corepluspenumbra\ estimation, respectively. This is despite the fact that the research community has been able to complete the puzzle and came up with methods, e.g., using \gls{cnn}s, that can incorporate the pathophysiology of \gls{ais} more naturally.
\bibliographystyle{plain}
\bibliography{references}
\end{document}

%% file: glossary.tex
\newglossaryentry{cbf}{name={CBF},description={cerebral blood flow}}
\newglossaryentry{cbv}{name={CBV},description={cerebral blood volume}}
\newglossaryentry{cpp}{name={CPP},description={cerebral perfusion pressure}}
\newglossaryentry{cvr}{name={CVR},description={cerebrovascular resistance}}
\newglossaryentry{oef}{name={OEF},description={oxygen extraction fraction}}
\newglossaryentry{ais}{name={AIS},description={acute ischemic stroke}}
\newglossaryentry{tso}{name={TsO},description={time since onset}}
\newglossaryentry{ttt}{name={TtT},description={time to treatment}}
\newglossaryentry{tici}{name={mTICI},description={modified Thrombolysis in Cerebral Infarction}}
\newglossaryentry{ivt}{name={IVT},description={intravenous treatment}}
\newglossaryentry{iat}{name={IAT},description={intra-arterial treatment}}
\newglossaryentry{evt}{name={EVT},description={endovascular treatment}}
\newglossaryentry{mrs}{name={mRS},description={modified Rankin Scale}}
\newglossaryentry{nihss}{name={NIHSS},description={National Institute of Health Stroke Scale}}
\newglossaryentry{ich}{name={ICH},description={intracranial hemorrhage}}
\newglossaryentry{ct}{name={CT},description={computed tomography}}
\newglossaryentry{ncct}{name={NCCT},description={non-enhanced or non-contrast CT}}
\newglossaryentry{aspects}{name={ASPECTS},description={Alberta Stroke Program Early CT Score}}
\newglossaryentry{mca}{name={MCA},description={middle cerebral artery}}
\newglossaryentry{mri}{name={MRI},description={magnetic resonance imaging}}
\newglossaryentry{dwi}{name={DWI},description={diffusion weighted imaging}}
\newglossaryentry{adc}{name={ADC},description={apparent diffusion coefficient}}
\newglossaryentry{ctp}{name={CTP},description={CT perfusion}}
\newglossaryentry{cta}{name={CTA},description={CT angiography}}
\newglossaryentry{pwi}{name={PWI},description={perfusion-weighted imaging}}
\newglossaryentry{tcc}{name={TCC},description={time concentration curve}}
\newglossaryentry{irf}{name={irf},description={impulse response function}}
\newglossaryentry{imp}{name={imp},description={Dirac impulse}}
\newglossaryentry{aif}{name={AIF},description={arterial imput function}}
\newglossaryentry{pet}{name={PET},description={positron emission tomography}}
\newglossaryentry{spect}{name={SPECT},description={single photon emission CT}}
\newglossaryentry{mra}{name={MRA},description={MRI angiography}}
\newglossaryentry{cc}{name={CC},description={collateral circulation}}
\newglossaryentry{3d}{name={3D},description={3 dimensional}}
\newglossaryentry{alara}{name={ALARA},description={as low as reasonably possible}}
\newglossaryentry{dcv}{name={DCV},description={deconvolution}}
\newglossaryentry{cnn}{name={CNN},description={convolutional neural network}}
\newglossaryentry{tmax}{name={Tmax},description={location of maximum of TCC\textsubscript{irf}}}
\newglossaryentry{rcbf}{name={rCBF},description={relative CBF}}
\newglossaryentry{cbct}{name={CBCT},description={C-arm or cone-beam CT}}
\newglossaryentry{cbctp}{name={CBCTP},description={CBCT perfusion}}
\newglossaryentry{de}{name={DE},description={dual-energy}}
\newglossaryentry{dencct}{name={DENCCT},description={DE NCCT}}
\newglossaryentry{rcbv}{name={rCBV},description={relative CBV}}
\newglossaryentry{o24h}{name={O24h},description={occlusion present at follow-up}}
\newglossaryentry{flair}{name={FLAIR},description={fluid-attenuated inversion recovery MRI}}
\newglossaryentry{vof}{name={VOF},description={venous output function}}
\newglossaryentry{glm}{name={GLM},description={generalized linear model}}
\newglossaryentry{csf}{name={CSF},description={cerebrospinal fluid}}
\newglossaryentry{minip}{name={minIP},description={minimum intensity projection}}
\newglossaryentry{meanip}{name={meanIP},description={mean intensity projection}}
\newglossaryentry{maxip}{name={maxIP},description={maximum intensity projection}}
\newglossaryentry{relu}{name={ReLU},description={rectified linear unit}}
\newglossaryentry{ica}{name={ICA},description={internal carotid artery}}
\newglossaryentry{aca}{name={ACA},description={arterior cerebral artery}}
\newglossaryentry{rep+}{name={Rep$^+$},description={(the group of) reperfusers}}
\newglossaryentry{rep-}{name={Rep$^-$},description={(the group of) non-reperfusers}}
\newglossaryentry{rep+-}{name={Rep$^{+/-}$},description={combination of Rep$^+$ and Rep$^-$}}
\newglossaryentry{lps}{name={LPS},description={left posterior superior}}
\newglossaryentry{mse}{name={MSE},description={mean squared error}}
\newglossaryentry{adv}{name={$|\Delta\mathrm{V}|$},description={absolute volume error}}
\newglossaryentry{rf}{name={RF},description={receptive field}}
\newglossaryentry{prf}{name={pRF},description={physical receptive field}}
\newglossaryentry{ce}{name={CE},description={cross-entropy}}
\newglossaryentry{sgd}{name={SGD},description={stochastic gradient descent}}
\newglossaryentry{dsc}{name={DSC},description={Dice coefficient}}
\newglossaryentry{dv}{name={$\Delta\mathrm{V}$},description={volume bias or error}}
\newglossaryentry{hd95}{name={HD95},description={95th percentile Hausdorff distance}}
\newglossaryentry{ppv}{name={PPV},description={precision or positive predictive value}}
\newglossaryentry{tpr}{name={TPR},description={recall or true positive rate}}
\newglossaryentry{ece}{name={ECE},description={expected calibration error}}
\newglossaryentry{auc}{name={AUC},description={area under the precision-recall curve}}
\newglossaryentry{fov}{name={FOV},description={field of view}}
\newglossaryentry{m}{name={M},description={MRCLEAN}}
\newglossaryentry{c}{name={C},description={CRISP}}
\newglossaryentry{k}{name={K},description={KAROLINSKA}}
\newglossaryentry{mck}{name={MCK},description={combination of M, C and K}}
\newglossaryentry{clpr}{name={ClPr},description={clinical practice}}
\newglossaryentry{jl}{name={JL},description={Julie Lambert}}
\newglossaryentry{jd}{name={JD},description={Jelle Demeestere}}
\newglossaryentry{grt}{name={GrT},description={grow time}}
\newglossaryentry{bao}{name={BAO},description={basilar artery occlusion}}
\newglossaryentry{rtpa}{name={rtPA},description={recombinant tissue plasminogen activator}}
\newglossaryentry{nn}{name={NN},description={neural network}}
\newglossaryentry{gpu}{name={GPU},description={graphics processing unit}}
\newglossaryentry{rgb}{name={RGB},description={red green blue}}
\newglossaryentry{pca}{name={PCA},description={principle component analysis}}
\newglossaryentry{fcn}{name={FCN},description={fully-convolutional network}}
\newglossaryentry{dl}{name={DL},description={Dice loss}}
\newglossaryentry{sd}{name={SD},description={soft Dice}}
\newglossaryentry{gan}{name={GAN},description={generative adversarial network}}
\newglossaryentry{t1w}{name={T1w},description={T1-weighted MRI}}
\newglossaryentry{stn}{name={STN},description={spatial transformer network}}
\newglossaryentry{dvn2}{name={DVN2},description={DeepVoxNet2}}
\newglossaryentry{gt}{name={GT},description={ground truth}}
\newglossaryentry{vs}{name={VS},description={voxel size}}
\newglossaryentry{lvo}{name={LVO},description={large vessel occlusion}}
\newglossaryentry{nexis}{name={NEXIS},description={NExt generation X-ray Imaging System}}
\newglossaryentry{ip}{name={IP},description={intensity projection}}

%% file: figures/vasculature.tex
\begin{figure}[t]
    \newcommand\myscale{\scaleppt}
    \setlength\tabcolsep{0pt}
    \centering
    \includegraphics[scale=\scale,scale=\myscale]{\rootpath 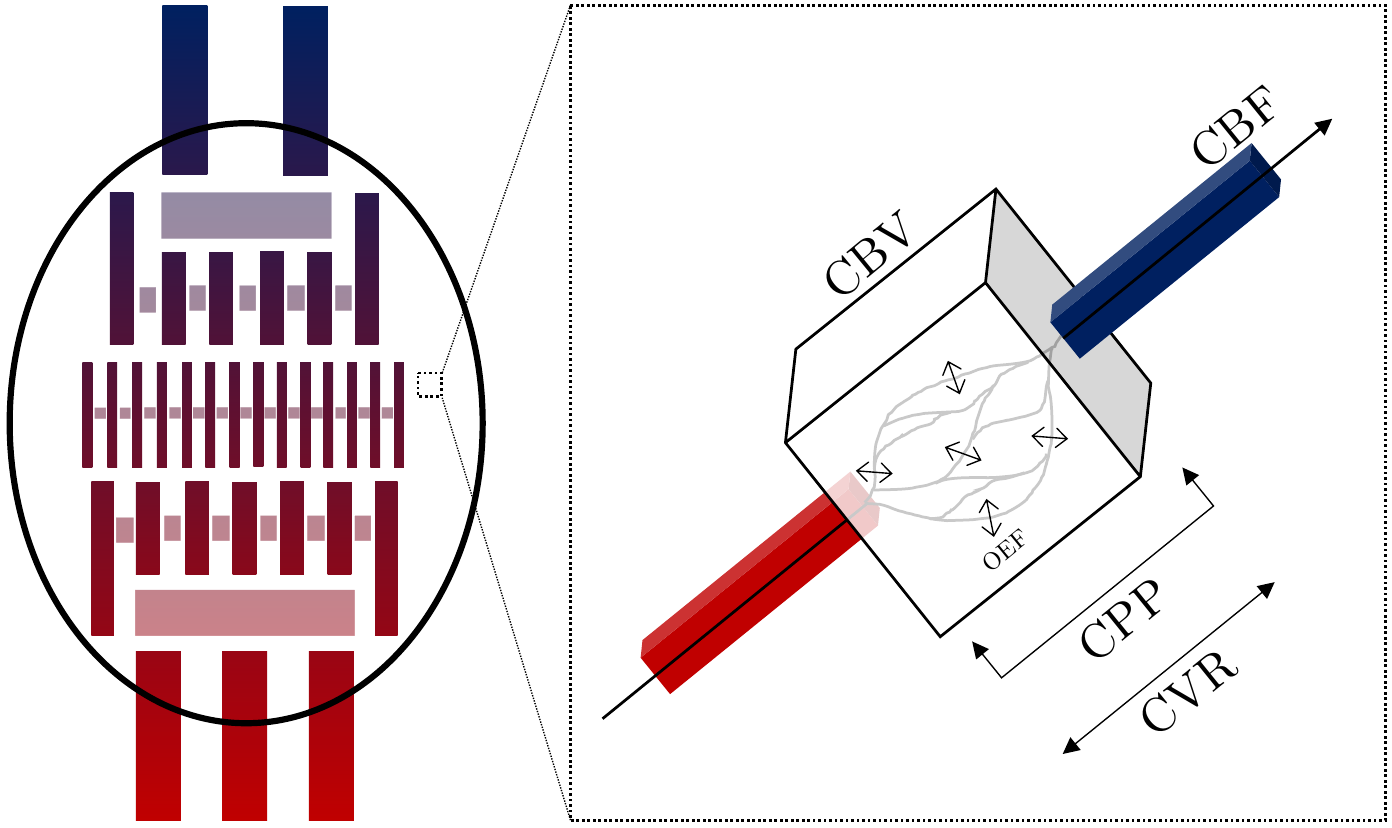}
    \caption[The cerebral vasculature.]{The cerebral vasculature. LEFT: A schematic representation of the cerebrovascular anatomy. Arterial and venous trees connect at the level of the capillaries, respectively carrying oxygen- and glucose-rich (RED) and -deprived (BLUE) blood. Transparent lateral connections refer to the collateral circulation, less or more pronounced across different subjects. RIGHT: A peace of tissue contains many capillaries and cells. Blood perfusion takes place when the \gls{cpp} overcomes the \gls{cvr}. A combination of the \gls{cbf}, \gls{cbv} and \gls{oef} summarizes the access to oxygen and glucose.}
    \label{fig:vasculature}
\end{figure}

%% file: figures/thresholds.tex
\begin{figure}[h]
    \newcommand\myscale{\scaleppt}
    \setlength\tabcolsep{0pt}
    \centering
    \includegraphics[scale=\scale,scale=\myscale]{\rootpath 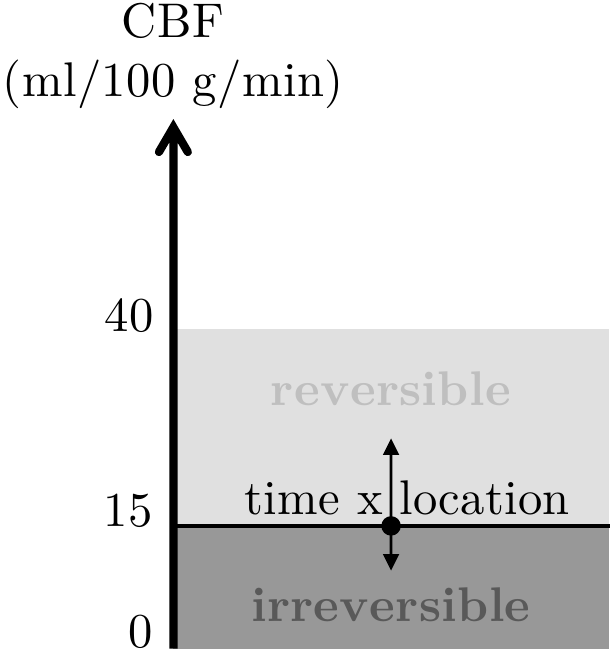}
    \caption[Important \gls{cbf} thresholds.]{During a pathological state of cerebral perfusion, the \gls{cbf} decreases progressively. It turns out that \gls{cbf} levels correlate with irreversible and reversible tissue damage. Nevertheless, absolute thresholds vary across different subjects, and with time and location due to a complex cascade of cellular reactions.}
    \label{fig:thresholds}
\end{figure}

%% file: figures/progression.tex
\begin{figure}[b!]
    \newcommand\myscale{\scaleppt}
    \setlength\tabcolsep{0pt}
    \centering
    \begin{tabular}{ccc}
        \multicolumn{3}{c}{\includegraphics[scale=\scale,scale=\myscale]{\rootpath 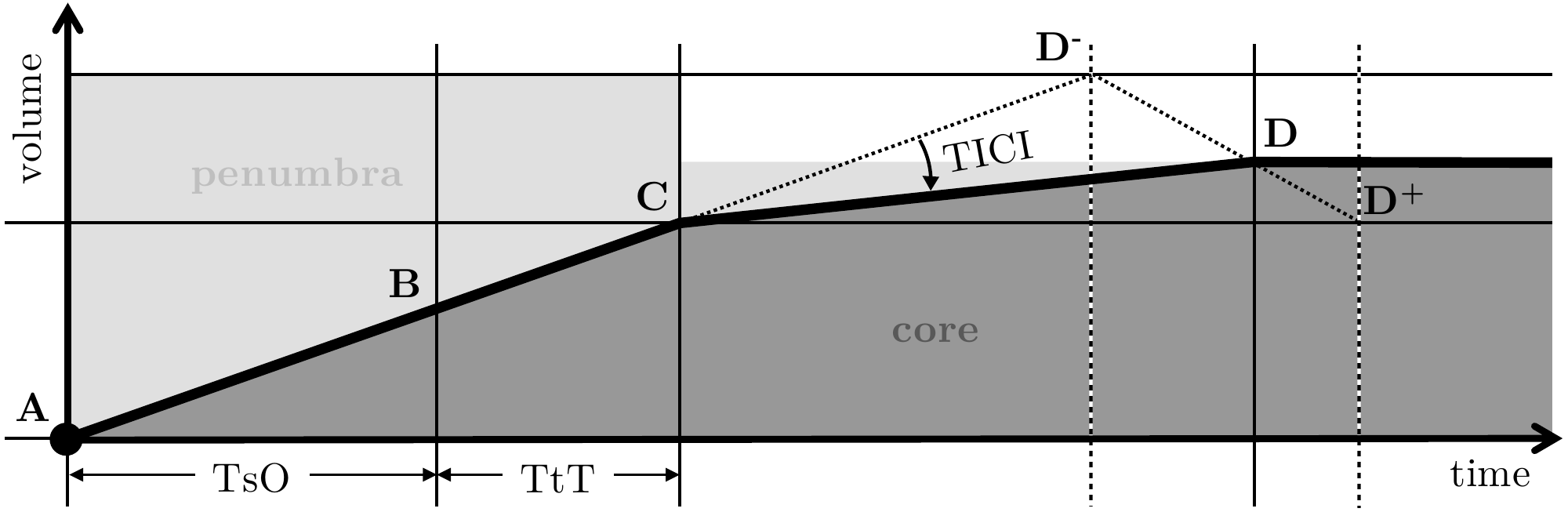}}
    \end{tabular}
    \begin{tabular}{ccc}
        \includegraphics[scale=\scale,scale=\myscale,trim={0 0 0 -0.5cm},clip]{\rootpath 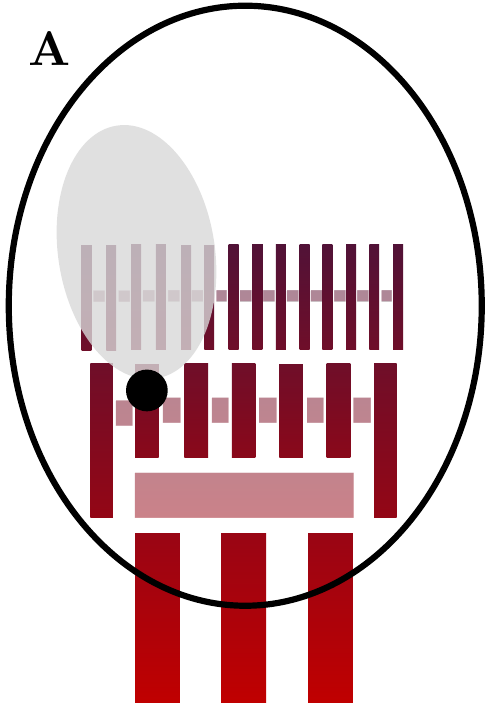}
        &\includegraphics[scale=\scale,scale=\myscale,trim={-1cm 0 -1cm -0.5cm},clip]{\rootpath 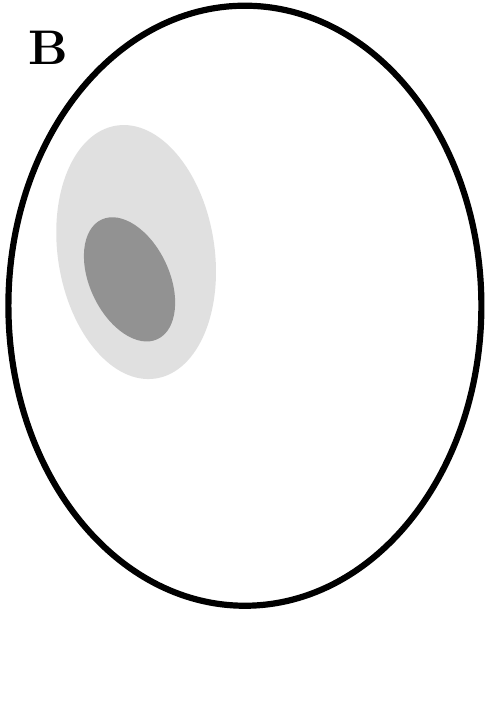}
        &\includegraphics[scale=\scale,scale=\myscale,trim={0 0 0 -0.5cm},clip]{\rootpath 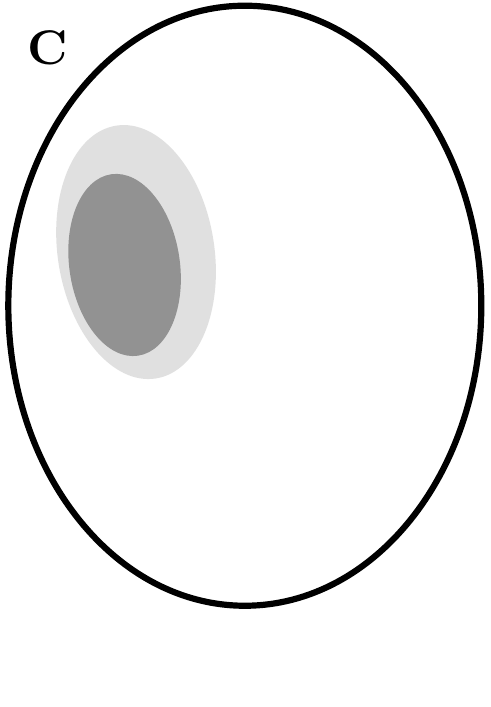}
    \end{tabular}\\
    \begin{tabular}{ccc}
        \includegraphics[scale=\scale,scale=\myscale,trim={0 1cm 0 0},clip]{\rootpath 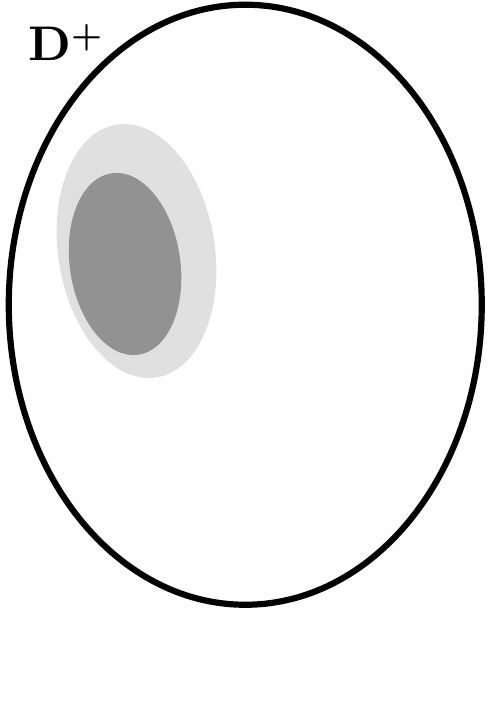}
        &\includegraphics[scale=\scale,scale=\myscale,trim={-1cm 1cm -1cm 0},clip]{\rootpath 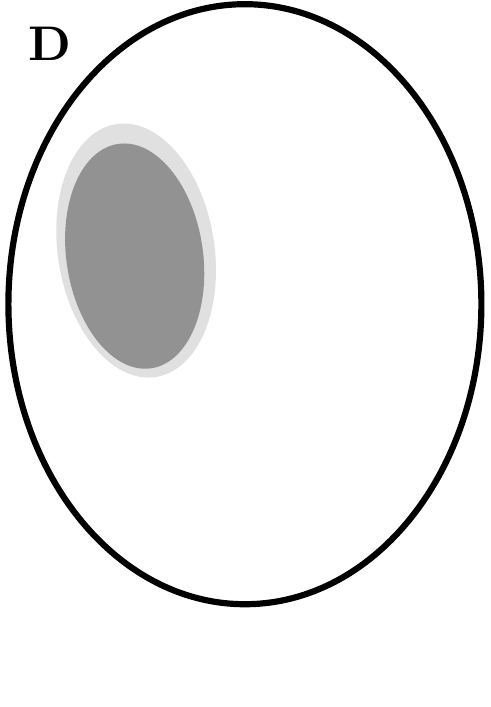}
        &\includegraphics[scale=\scale,scale=\myscale,trim={0 1cm 0 0},clip]{\rootpath 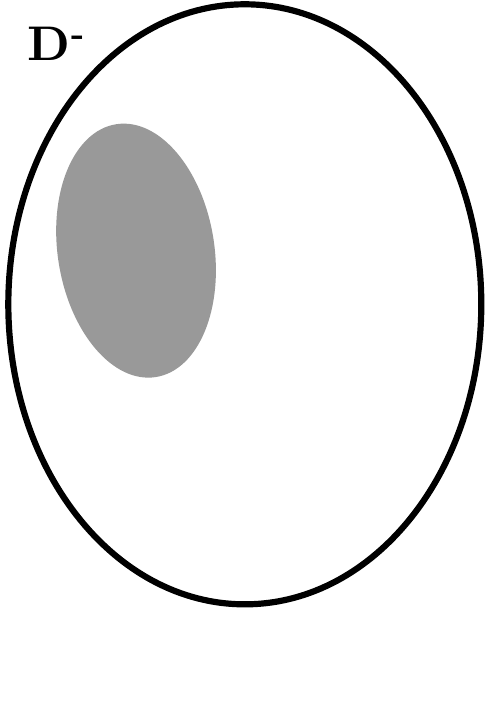} 
    \end{tabular}
    \caption[Progression of core and penumbra in \gls{ais}.]{At the onset of an \gls{ais} there is no irreversible damage (A). After a certain \gls{tso}, the patient is triaged for a certain treatment, a.o. based on acute imaging and estimates of the tissue state (i) at that moment (B), and (ii) at the moment when the (partial) reperfusion takes place, i.e. after a certain \gls{ttt} (C). The treatment affects the extent of the final infarction (D- > D > D+) and may slow down the progression (decreasing the tissue hypoperfusion will suppress its malicious effect on surrounding tissue). To assess the final infarction, follow-up imaging should be acquired not sooner than when the infarction has fully progressed (D+).}
    \label{fig:progression}
\end{figure}

%% file: figures/deconvolution.tex
\begin{figure}[b]
    \newcommand\myscale{\scaleppt} 
    \setlength\tabcolsep{0pt}
    \centering
    \includegraphics[scale=\scale,scale=\myscale]{\rootpath 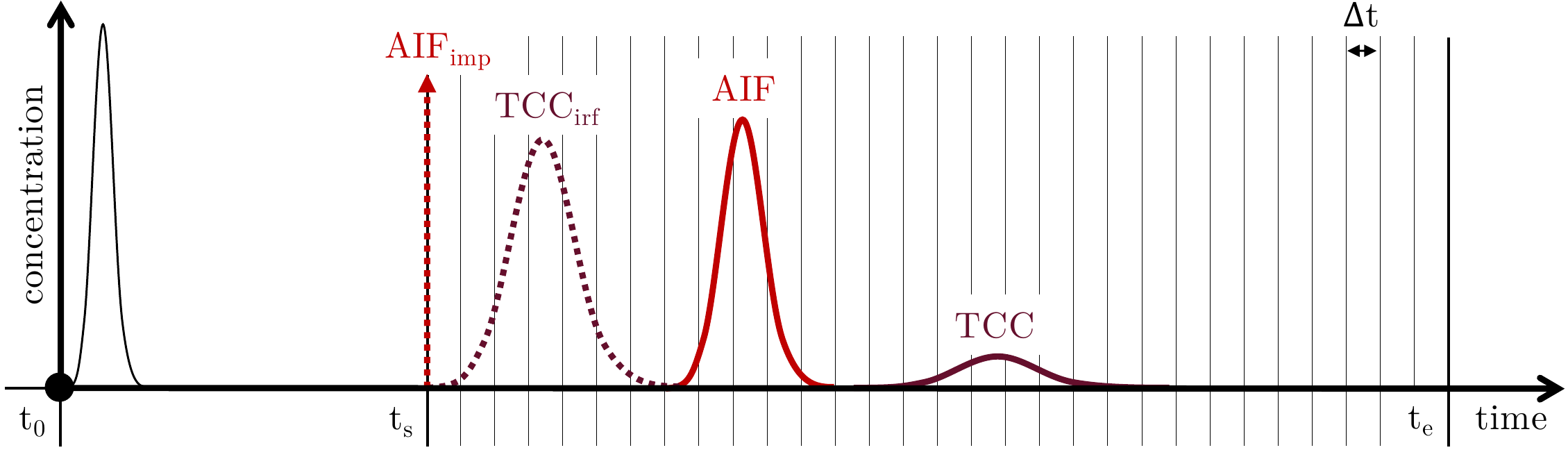}
    \caption[Perfusion imaging and deconvolution.]{Perfusion imaging acquires a series of images between t\textsubscript{s} and t\textsubscript{e} after the injection of a contrast agent at t\textsubscript{0}. The measured \gls{tcc}s will look different across different patients and measurements, even so for the injection location itself and for the location of the \gls{aif} at the entrance of the brain. Measurements can be normalized, i.e. \gls{tcc}\textsubscript{\gls{irf}}, by deconvolution, as if they were the result of an \gls{aif}\textsubscript{\gls{imp}}. Straightforwardly, the timing of the start t\textsubscript{s}-t\textsubscript{0} and end t\textsubscript{e}-t\textsubscript{0} of the scan is of crucial importance. The time resolution $\Delta$t is important for the deconvolution to be mathematically valid. From the resulting \gls{tcc}\textsubscript{\gls{irf}}s certain perfusion dynamics can be derived directly.}
    \label{fig:deconvolution}
\end{figure}

%% file: figures/machine_learning.tex
\begin{figure}[h]
    \newcommand\myscale{\scaleppt} 
    \setlength\tabcolsep{0pt}
    \centering
    \includegraphics[scale=\scale,scale=\myscale]{\rootpath 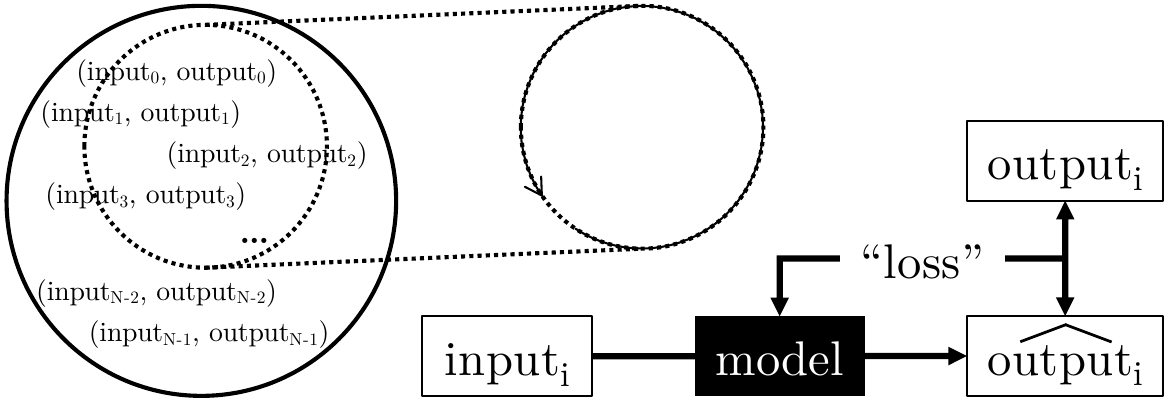}
    \caption[A schematic representation of data-driven methods in \gls{ais}.]{Data-driven methods for automated tissue status segmentation in \gls{ais} need (i) a dataset of input-output pairs, and (ii) a model that maps an input to an output. The model has trainable parameters that are tuned (iteratively) to minimize the loss between the observed and predicted output. Current methods mostly provide models for different subsets of the original dataset, e.g. one for core and one for penumbra estimation, respectively using a dataset of reperfusers and non-reperfusers.}
    \label{fig:machine_learning}
\end{figure}

%% file: template.bbl
\begin{thebibliography}{100}

\bibitem{AguiardeSousa2019}
Diana {Aguiar de Sousa}, Rascha von Martial, S{\`{o}}nia Abilleira, ..., Valery
  Feigin, Valeria Caso, and Urs Fischer.
\newblock {Access to and delivery of acute ischaemic stroke treatments: A
  survey of national scientific societies and stroke experts in 44 European
  countries}.
\newblock {\em European Stroke Journal}, 4(1):13--28, March 2019.

\bibitem{Albers2018}
Gregory~W. Albers, Michael~P. Marks, Stephanie Kemp, ..., Philip~W Lavori,
  Joseph~P Broderick, and Maarten~G Lansberg.
\newblock {Thrombectomy for Stroke at 6 to 16 Hours with Selection by Perfusion
  Imaging}.
\newblock {\em New England Journal of Medicine}, 378(8):708--718, February
  2018.

\bibitem{Albers2006}
Gregory~W. Albers, Vincent~N. Thijs, Lawrence Wechsler, ..., Scott Hamilton,
  Michael Moseley, and Michael~P. Marks.
\newblock {Magnetic resonance imaging profiles predict clinical response to
  early reperfusion: The diffusion and perfusion imaging evaluation for
  understanding stroke evolution (DEFUSE) study}.
\newblock {\em Annals of Neurology}, 60(5):508--517, November 2006.

\bibitem{Astrup1981}
J~Astrup, B~K Siesj{\"{o}}, and L~Symon.
\newblock {Thresholds in cerebral ischemia - the ischemic penumbra.}
\newblock {\em Stroke}, 12(6):723--725, November 1981.

\bibitem{Bang2011}
Oh~Young Bang, Jeffrey~L. Saver, Suk~Jae Kim, ..., Bruce Ovbiagele, Kwang~Ho
  Lee, and David~S. Liebeskind.
\newblock {Collateral Flow Predicts Response to Endovascular Therapy for Acute
  Ischemic Stroke}.
\newblock {\em Stroke}, 42(3):693--699, March 2011.

\bibitem{Barber2000}
Philip~A Barber, Andrew~M Demchuk, Jinjin Zhang, and Alastair~M Buchan.
\newblock {Validity and reliability of a quantitative computed tomography score
  in predicting outcome of hyperacute stroke before thrombolytic therapy}.
\newblock {\em The Lancet}, 355(9216):1670--1674, May 2000.

\bibitem{Baron1981}
J.C. Baron, M.G. Bousser, D.~Comar, F.~Soussaline, and P.~Castaigne.
\newblock {Noninvasive Tomographic Study of Cerebral Blood Flow and Oxygen
  Metabolism in vivo}.
\newblock {\em European Neurology}, 20(3):273--284, 1981.

\bibitem{Baron1999}
Jean-Claude Baron.
\newblock {Mapping the Ischaemic Penumbra with PET: Implications for Acute
  Stroke Treatment}.
\newblock {\em Cerebrovascular Diseases}, 9(4):193--201, 1999.

\bibitem{Berkhemer2015}
Olvert~A. Berkhemer, Puck~S.S. Fransen, Debbie Beumer, ..., Robert~J. van
  Oostenbrugge, Charles~B.L.M. Majoie, and Diederik~W.J. Dippel.
\newblock {A Randomized Trial of Intraarterial Treatment for Acute Ischemic
  Stroke}.
\newblock {\em New England Journal of Medicine}, 372(1):11--20, January 2015.

\bibitem{Bertels2019a}
Jeroen Bertels, David Robben, Dirk Vandermeulen, and Paul Suetens.
\newblock {Contra-Lateral Information CNN for Core Lesion Segmentation Based on
  Native CTP in Acute Stroke}.
\newblock In {\em LNCS}, volume 11383, pages 263--270. Springer Nature, 2019.

\bibitem{Bivard2013}
Andrew Bivard, Christopher Levi, Neil Spratt, and Mark Parsons.
\newblock {Perfusion CT in Acute Stroke: A Comprehensive Analysis of Infarct
  and Penumbra}.
\newblock {\em Radiology}, 267(2):543--550, May 2013.

\bibitem{Boutelier2012}
Timoth{\'{e}} Boutelier, Koshuke Kudo, Fabrice Pautot, and Makoto Sasaki.
\newblock {Bayesian Hemodynamic Parameter Estimation by Bolus Tracking
  Perfusion Weighted Imaging}.
\newblock {\em IEEE Transactions on Medical Imaging}, 31(7):1381--1395, July
  2012.

\bibitem{Campbell2019}
Bruce C~V Campbell, Henry Ma, Peter~A Ringleb, ..., R.~Burns, C.~Johnston, and
  M.~Williams.
\newblock {Extending thrombolysis to 4{\textperiodcentered}5–9 h and wake-up
  stroke using perfusion imaging: a systematic review and meta-analysis of
  individual patient data}.
\newblock {\em The Lancet}, 394(10193):139--147, July 2019.

\bibitem{Campbell2011}
Bruce~C.V. Campbell, S{\o}ren Christensen, Christopher~R. Levi, ...,
  Geoffrey~A. Donnan, Stephen~M. Davis, and Mark~W. Parsons.
\newblock {Cerebral blood flow is the optimal CT perfusion parameter for
  assessing infarct core}.
\newblock {\em Stroke}, 42(12):3435--3440, 2011.

\bibitem{Campbell2015}
Bruce~C.V. Campbell, Peter~J. Mitchell, Timothy~J. Kleinig, ..., Patricia~M.
  Desmond, Geoffrey~A. Donnan, and Stephen~M. Davis.
\newblock {Endovascular Therapy for Ischemic Stroke with Perfusion-Imaging
  Selection}.
\newblock {\em New England Journal of Medicine}, 372(11):1009--1018, March
  2015.

\bibitem{Carroll2008}
Timothy~J. Carroll, Sandra Horowitz, Wanyong Shin, ..., Saad Ali, Jeffrey
  Raizer, and Stephen Futterer.
\newblock {Quantification of cerebral perfusion using the “bookend
  technique”: an evaluation in CNS tumors}.
\newblock {\em Magnetic Resonance Imaging}, 26(10):1352--1359, December 2008.

\bibitem{Castillo1997}
Jos{\'{e}} Castillo, Antoni D{\'{a}}valos, and Manuel Noya.
\newblock {Progression of ischaemic stroke and excitotoxic aminoacids}.
\newblock {\em The Lancet}, 349(9045):79--82, January 1997.

\bibitem{Cereda2016}
Carlo~W. Cereda, S{\o}ren Christensen, Bruce~C.V. Campbell, ..., Gregory~W.
  Albers, Mark~W. Parsons, and Maarten~G. Lansberg.
\newblock {A benchmarking tool to evaluate computer tomography perfusion
  infarct core predictions against a DWI standard}.
\newblock {\em Journal of Cerebral Blood Flow {\&} Metabolism},
  36(10):1780--1789, October 2016.

\bibitem{Chen2019}
Yu~Chen, Yuexiang Li, and Yefeng Zheng.
\newblock {Ensembles of Modalities Fused Model for Ischemic Stroke Lesion
  Segmentation}.
\newblock In {\em LNCS}. Springer, 2019.

\bibitem{Christoforidis2009}
G.A. Christoforidis, C.~Karakasis, Y.~Mohammad, L.P. Caragine, M.~Yang, and
  A.P. Slivka.
\newblock {Predictors of Hemorrhage Following Intra-Arterial Thrombolysis for
  Acute Ischemic Stroke: The Role of Pial Collateral Formation}.
\newblock {\em American Journal of Neuroradiology}, 30(1):165--170, January
  2009.

\bibitem{Coutts2005}
Shelagh~B. Coutts, Jessica~E. Simon, Michael Eliasziw, ..., James~N. Scott,
  Alastair~M. Buchan, and Andrew~M. Demchuk.
\newblock {Triaging transient ischemic attack and minor stroke patients using
  acute magnetic resonance imaging}.
\newblock {\em Annals of Neurology}, 57(6):848--854, June 2005.

\bibitem{DelaRosa2020}
Ezequiel de~la Rosa, David Robben, Diana~M. Sima, Jan~S. Kirschke, and Bjoern
  Menze.
\newblock {Differentiable Deconvolution for Improved Stroke Perfusion
  Analysis}.
\newblock In {\em LNCS: MICCAI 2020}, pages 593--602. Springer International
  Publishing, 2020.

\bibitem{DelZoppo2000}
Gregory~J del Zoppo and John~M Hallenbeck.
\newblock {Advances in the Vascular Pathophysiology of Ischemic Stroke}.
\newblock {\em Thrombosis Research}, 98(3):73--81, May 2000.

\bibitem{DEsterre2015}
Christopher~D. D'Esterre, Mari~E. Boesen, Seong~Hwan Ahn, ..., Mayank Goyal,
  Ting~Y. Lee, and Bijoy~K. Menon.
\newblock {Time-Dependent Computed Tomographic Perfusion Thresholds for
  Patients With Acute Ischemic Stroke}.
\newblock {\em Stroke}, 46(12):3390--3397, December 2015.

\bibitem{Fahmi2012}
F.~Fahmi, H.A. Marquering, G.J. Streekstra, ..., B.K. Velthuis, E.~VanBavel,
  and C.B. Majoie.
\newblock {Differences in CT Perfusion Summary Maps for Patients with Acute
  Ischemic Stroke Generated by 2 Software Packages}.
\newblock {\em American Journal of Neuroradiology}, 33(11):2074--2080, December
  2012.

\bibitem{Farzin2016}
Behzad Farzin, Robert Fahed, Francois Guilbert, ..., Marc-Antoine Henry, Tim~E
  Darsaut, and Jean Raymond.
\newblock {Early CT changes in patients admitted for thrombectomy}.
\newblock {\em Neurology}, 87(3):249--256, July 2016.

\bibitem{Fieselmann2011}
Andreas Fieselmann, Markus Kowarschik, Arundhuti Ganguly, Joachim Hornegger,
  and Rebecca Fahrig.
\newblock {Deconvolution-Based CT and MR Brain Perfusion Measurement:
  Theoretical Model Revisited and Practical Implementation Details}.
\newblock {\em International Journal of Biomedical Imaging}, 2011:1--20, 2011.

\bibitem{Flores2015}
Alan Flores, Marta Rubiera, Marc Rib{\'{o}}, ..., Miguel Lemus, Pilar
  Coscojuela, and Carlos~A. Molina.
\newblock {Poor Collateral Circulation Assessed by Multiphase Computed
  Tomographic Angiography Predicts Malignant Middle Cerebral Artery Evolution
  After Reperfusion Therapies}.
\newblock {\em Stroke}, 46(11):3149--3153, November 2015.

\bibitem{Fugate2015}
Jennifer~E. Fugate and Alejandro~A. Rabinstein.
\newblock {Absolute and Relative Contraindications to IV rt-PA for Acute
  Ischemic Stroke}.
\newblock {\em The Neurohospitalist}, 5(3):110--121, July 2015.

\bibitem{Garcia1984}
J~H Garcia.
\newblock {Experimental ischemic stroke: a review.}
\newblock {\em Stroke}, 15(1):5--14, January 1984.

\bibitem{Garcia1988}
J~H Garcia.
\newblock {Morphology of global cerebral ischemia}.
\newblock {\em Critical Care Medicine}, 16(10):979--987, October 1988.

\bibitem{Garcia1983}
J~H Garcia, J~B Hazelrig, H~L Mitchem, ..., A~G Hudetz, J~H Halsey, and K~A
  Conger.
\newblock {Transient Focal Ischemia in Subhuman Primates}.
\newblock {\em Journal of Neuropathology and Experimental Neurology},
  42(1):44--60, January 1983.

\bibitem{Garcia1993}
J~H Garcia, Y~Yoshida, H~Chen, ..., J~Lian, S~Chen, and M~Chopp.
\newblock {Progression from ischemic injury to infarct following middle
  cerebral artery occlusion in the rat.}
\newblock {\em The American journal of pathology}, 142(2):623--35, February
  1993.

\bibitem{Garcia1995}
Julio~H. Garcia, Kai-Feng Liu, and Khang-Loon Ho.
\newblock {Neuronal Necrosis After Middle Cerebral Artery Occlusion in Wistar
  Rats Progresses at Different Time Intervals in the Caudoputamen and the
  Cortex}.
\newblock {\em Stroke}, 26(4):636--643, April 1995.

\bibitem{Garcia1997}
Julio~H. Garcia, Kai-Feng Liu, Zhu-Rong Ye, and Jorge~A. Gutierrez.
\newblock {Incomplete Infarct and Delayed Neuronal Death After Transient Middle
  Cerebral Artery Occlusion in Rats}.
\newblock {\em Stroke}, 28(11):2303--2310, November 1997.

\bibitem{Goyal2015}
Mayank Goyal, Andrew~M. Demchuk, Bijoy~K. Menon, ..., Mark~W. Lowerison,
  Tolulope~T. Sajobi, and Michael~D. Hill.
\newblock {Randomized Assessment of Rapid Endovascular Treatment of Ischemic
  Stroke}.
\newblock {\em New England Journal of Medicine}, 372(11):1019--1030, March
  2015.

\bibitem{Goyal2016}
Mayank Goyal, Bijoy~K Menon, Wim~H van Zwam, ..., Jeffrey~L Saver, Michael~D
  Hill, and Tudor~G Jovin.
\newblock {Endovascular thrombectomy after large-vessel ischaemic stroke: a
  meta-analysis of individual patient data from five randomised trials}.
\newblock {\em The Lancet}, 387(10029):1723--1731, April 2016.

\bibitem{Goyal2020}
Mayank Goyal, Johanna~M. Ospel, Bijoy Menon, ..., Michael~D. Hill, Diederik
  Dippel, and Marc Fisher.
\newblock {Challenging the Ischemic Core Concept in Acute Ischemic Stroke
  Imaging}.
\newblock {\em Stroke}, 51(10):3147--3155, October 2020.

\bibitem{Guenego2018}
Adrien Guenego, Michael Mlynash, Soren Christensen, ..., Jeremy~J. Heit,
  Maarten~G. Lansberg, and Gregory~W. Albers.
\newblock {Hypoperfusion ratio predicts infarct growth during transfer for
  thrombectomy}.
\newblock {\em Annals of Neurology}, 84(4):616--620, October 2018.

\bibitem{Hacke2008}
Werner Hacke, Markku Kaste, Erich Bluhmki, ..., R{\"{u}}diger von Kummer, Nils
  Wahlgren, and Danilo Toni.
\newblock {Thrombolysis with Alteplase 3 to 4.5 Hours after Acute Ischemic
  Stroke}.
\newblock {\em New England Journal of Medicine}, 2008.

\bibitem{Hampton-till2015}
James Hampton-Till, Michael Harrison, Anna~Luisa K{\"{u}}hn, ..., Eric
  Greveson, Michalis Papadakis, and Iris~Q Grunwald.
\newblock {Automated Quantification of Stroke Damage on Brain Computed
  Tomography Scans: e-ASPECTS}.
\newblock {\em European Medical Journal of Neurology}, 3(1):69--74, 2015.

\bibitem{Heiss1983}
W.-D. Heiss and G.~Rosner.
\newblock {Functional recovery of cortical neurons as related to degree and
  duration of ischemia}.
\newblock {\em Annals of Neurology}, 14(3):294--301, September 1983.

\bibitem{Herweh2016}
Christian Herweh, Peter~A. Ringleb, Geraldine Rauch, ..., Daniel Richter, Simon
  Schieber, and Simon Nagel.
\newblock {Performance of e-ASPECTS software in comparison to that of stroke
  physicians on assessing CT scans of acute ischemic stroke patients}.
\newblock {\em International Journal of Stroke}, 11(4):438--445, June 2016.

\bibitem{Hess2018}
Andreas Hess, Raphael Meier, Johannes Kaesmacher, ..., David Liebeskind, Roland
  Wiest, and Richard McKinley.
\newblock {Synthetic Perfusion Maps: Imaging Perfusion Deficits in DSC-MRI with
  Deep Learning}.
\newblock In {\em LNCS}, pages 447--455. Springer, 2019.

\bibitem{Higashida2003}
Randall~T. Higashida and Anthony~J. Furlan.
\newblock {Trial Design and Reporting Standards for Intra-Arterial Cerebral
  Thrombolysis for Acute Ischemic Stroke}.
\newblock {\em Stroke}, 34(8), August 2003.

\bibitem{Ho2016}
King~Chung Ho, Fabien Scalzo, Karthik~V. Sarma, Suzie El-Saden, and Corey~W.
  Arnold.
\newblock {A temporal deep learning approach for MR perfusion parameter
  estimation in stroke}.
\newblock In {\em 2016 23rd International Conference on Pattern Recognition
  (ICPR)}, pages 1315--1320. IEEE, December 2016.

\bibitem{Hossmann1994}
K-A. Hossmann.
\newblock {Viability thresholds and the penumbra of focal ischemia}.
\newblock {\em Annals of Neurology}, 36(4):557--565, October 1994.

\bibitem{Hossmann1996}
K~A Hossmann.
\newblock {Periinfarct depolarizations.}
\newblock {\em Cerebrovascular and brain metabolism reviews}, 8(3):195--208,
  1996.

\bibitem{Hossmann2006}
Konstantin-Alexander Hossmann.
\newblock {Pathophysiology and Therapy of Experimental Stroke}.
\newblock {\em Cellular and Molecular Neurobiology}, 26(7-8):1055--1081,
  December 2006.

\bibitem{Ioannidis2021}
Georgios~S. Ioannidis, S{\o}ren Christensen, Katerina Nikiforaki, ..., Mauricio
  Reyes, Maarten Lansberg, and Kostas Marias.
\newblock {Cerebral CT Perfusion in Acute Stroke: The Effect of Lowering the
  Tube Load and Sampling Rate on the Reproducibility of Parametric Maps}.
\newblock {\em Diagnostics}, 11(6):1121, June 2021.

\bibitem{Ito1979}
U~Ito, K~Ohno, R~Nakamura, F~Suganuma, and Y~Inaba.
\newblock {Brain edema during ischemia and after restoration of blood flow.
  Measurement of water, sodium, potassium content and plasma protein
  permeability.}
\newblock {\em Stroke}, 10(5):542--547, September 1979.

\bibitem{Jones1981}
Thomas~H. Jones, Richard~B. Morawetz, Robert~M. Crowell, ..., Stuart~J.
  FitzGibbon, Umberto DeGirolami, and Robert~G. Ojemann.
\newblock {Thresholds of focal cerebral ischemia in awake monkeys}.
\newblock {\em Journal of Neurosurgery}, 54(6):773--782, June 1981.

\bibitem{Jovin2015}
Tudor~G. Jovin, Angel Chamorro, Erik Cobo, ..., R{\"{u}}diger von Kummer,
  Miquel Gallofr{\'{e}}, and Antoni D{\'{a}}valos.
\newblock {Thrombectomy within 8 Hours after Symptom Onset in Ischemic Stroke}.
\newblock {\em New England Journal of Medicine}, 372(24):2296--2306, June 2015.

\bibitem{Kemmling2015}
Andr{\'{e}} Kemmling, Fabian Flottmann, Nils~Daniel Forkert, ..., Marios
  Psychogios, Soenke Langner, and Jens Fiehler.
\newblock {Multivariate Dynamic Prediction of Ischemic Infarction and Tissue
  Salvage as a Function of Time and Degree of Recanalization}.
\newblock {\em Journal of Cerebral Blood Flow {\&} Metabolism},
  35(9):1397--1405, September 2015.

\bibitem{Kety1950}
Seymour~S. Kety.
\newblock {Circulation and metabolism of the human brain in health and
  disease}.
\newblock {\em The American Journal of Medicine}, 8(2):205--217, February 1950.

\bibitem{Koga2020}
Masatoshi Koga, Haruko Yamamoto, Manabu Inoue, ..., Kazumi Kimura, Kazuo
  Minematsu, and Kazunori Toyoda.
\newblock {Thrombolysis With Alteplase at 0.6 mg/kg for Stroke With Unknown
  Time of Onset}.
\newblock {\em Stroke}, 51(5):1530--1538, May 2020.

\bibitem{Kuang2019}
H.~Kuang, M.~Najm, D.~Chakraborty, ..., A.M. Demchuk, B.K. Menon, and Wu~Qiu.
\newblock {Automated ASPECTS on Noncontrast CT Scans in Patients with Acute
  Ischemic Stroke Using Machine Learning}.
\newblock {\em American Journal of Neuroradiology}, 40(1):33--38, January 2019.

\bibitem{Lee2008a}
T.-J. Lee and B.~Murphy.
\newblock {Implementing Deconvolution Analysis for Perfusion CT}.
\newblock In {\em Multidetector Computed Tomography in Cerebrovascular Disease:
  CT Perfusion Imaging}, volume~29, pages e18--e19. Informa Healthcare, 4
  edition, 2008.

\bibitem{Liebeskind2003}
David~S. Liebeskind.
\newblock {Collateral Circulation}.
\newblock {\em Stroke}, 34(9):2279--2284, September 2003.

\bibitem{Lin2016}
Longting Lin, Andrew Bivard, Venkatesh Krishnamurthy, Christopher~R Levi, and
  Mark~W Parsons.
\newblock {Whole-Brain CT Perfusion to Quantify Acute Ischemic Penumbra and
  Core}.
\newblock {\em Radiology}, 279(3):876--887, June 2016.

\bibitem{Lin2003}
Weili Lin, Jin-Moo Lee, Yueh~Z. Lee, Katie~D. Vo, Thomas Pilgram, and Chung~Y.
  Hsu.
\newblock {Temporal Relationship Between Apparent Diffusion Coefficient and
  Absolute Measurements of Cerebral Blood Flow in Acute Stroke Patients}.
\newblock {\em Stroke}, 34(1):64--70, January 2003.

\bibitem{Liu2019b}
Pengbo Liu.
\newblock {Stroke Lesion Segmentation with 2D Novel CNN Pipeline and Novel Loss
  Function}.
\newblock In {\em LNCS}, volume 11383, pages 253--262. Springer, 2019.

\bibitem{Lorenz2006}
Cory Lorenz, Thomas Benner, Poe~Jou Chen, ..., Yawu Liu, Juho Nuutinen, and
  A.~Gregory Sorensen.
\newblock {Automated perfusion-weighted MRI using localized arterial input
  functions}.
\newblock {\em Journal of Magnetic Resonance Imaging}, 24:1133--1139, March
  2006.

\bibitem{Lucas2018}
Christian Lucas, Andr{\'{e}} Kemmling, Nassim Bouteldja, Linda~F. Aulmann, Amir
  {Madany Mamlouk}, and Mattias~P. Heinrich.
\newblock {Learning to Predict Ischemic Stroke Growth on Acute CT Perfusion
  Data by Interpolating Low-Dimensional Shape Representations}.
\newblock {\em Frontiers in Neurology}, 9(NOV):1--15, November 2018.

\bibitem{Ma2019}
Henry Ma, Bruce~C.V. Campbell, Mark~W. Parsons, ..., Atte Meretoja, Stephen~M.
  Davis, and Geoffrey~A. Donnan.
\newblock {Thrombolysis Guided by Perfusion Imaging up to 9 Hours after Onset
  of Stroke}.
\newblock {\em New England Journal of Medicine}, 380(19):1795--1803, May 2019.

\bibitem{Maas2009}
Matthew~B. Maas, Michael~H. Lev, Hakan Ay, ..., Andre Kemmling, Walter~J.
  Koroshetz, and Karen~L. Furie.
\newblock {Collateral Vessels on CT Angiography Predict Outcome in Acute
  Ischemic Stroke}.
\newblock {\em Stroke}, 40(9):3001--3005, September 2009.

\bibitem{Marchal1996}
Gilles Marchal, Vincent Beaudouin, Patrice Rioux, ..., Fausto Viader,
  Jean~Michel Derlon, and Jean~Claude Baron.
\newblock {Prolonged Persistence of Substantial Volumes of Potentially Viable
  Brain Tissue After Stroke}.
\newblock {\em Stroke}, 27(4):599--606, April 1996.

\bibitem{Markus2004}
H.~S. Markus.
\newblock {Cerebral perfusion and stroke}.
\newblock {\em Journal of Neurology, Neurosurgery {\&} Psychiatry},
  75(3):353--361, March 2004.

\bibitem{Marler1995}
J.~R. Marler, T.~Brott, P.~Lyden, ..., E.~C. Haley, M.~Meyer, and B.~C. Tilley.
\newblock {Tissue Plasminogen Activator for Acute Ischemic Stroke}.
\newblock {\em New England Journal of Medicine}, 333(24):1581--1588, December
  1995.

\bibitem{McKinley2017}
Richard McKinley, Levin H{\"{a}}ni, Jan Gralla, ..., Kaspar Mattmann, Mauricio
  Reyes, and Roland Wiest.
\newblock {Fully automated stroke tissue estimation using random forest
  classifiers (FASTER)}.
\newblock {\em Journal of Cerebral Blood Flow {\&} Metabolism},
  37(8):2728--2741, August 2017.

\bibitem{McTaggart2015}
Ryan~A. McTaggart, Tudor~G. Jovin, Maarten~G. Lansberg, ..., Manabu Inoue,
  Michael~P. Marks, and Gregory~W. Albers.
\newblock {Alberta Stroke Program Early Computed Tomographic Scoring
  Performance in a Series of Patients Undergoing Computed Tomography and MRI}.
\newblock {\em Stroke}, 46(2):407--412, February 2015.

\bibitem{Meijs2016}
Midas Meijs, Soren Christensen, Maarten~G. Lansberg, Gregory~W. Albers, and
  Fernando Calamante.
\newblock {Analysis of perfusion MRI in stroke: To deconvolve, or not to
  deconvolve}.
\newblock {\em Magnetic Resonance in Medicine}, 76(4):1282--1290, October 2016.

\bibitem{Menon2015a}
Bijoy~K. Menon, Christopher~D. D'Esterre, Emmad~M. Qazi, ..., Leszek Hahn,
  Andrew~M. Demchuk, and Mayank Goyal.
\newblock {Multiphase CT Angiography: A New Tool for the Imaging Triage of
  Patients with Acute Ischemic Stroke}.
\newblock {\em Radiology}, 275(2):510--520, May 2015.

\bibitem{Menon2013a}
Bijoy~K Menon, Billy O'Brien, Andrew Bivard, ..., Xuewen Lu, Christopher Levi,
  and Mark~W Parsons.
\newblock {Assessment of Leptomeningeal Collaterals Using Dynamic CT
  Angiography in Patients with Acute Ischemic Stroke}.
\newblock {\em Journal of Cerebral Blood Flow {\&} Metabolism}, 33(3):365--371,
  March 2013.

\bibitem{Menon2013}
Bijoy~K. Menon, Eric~E. Smith, Shelagh~B. Coutts, ..., Hyuk-Won Chang, Jeong-Ho
  Hong, and Sung~Il Sohn.
\newblock {Leptomeningeal collaterals are associated with modifiable metabolic
  risk factors}.
\newblock {\em Annals of Neurology}, 74(2):241--248, August 2013.

\bibitem{Mraovitch1996}
Sima Mraovitch and Richard Sercombe.
\newblock {\em {Neurophysiological Basis of Cerebral Blood Flow Control: An
  Introduction}}.
\newblock John Libbey {\&} Co., 1 edition, 1996.

\bibitem{Nagel2017}
Simon Nagel, Devesh Sinha, Diana Day, ..., James Hampton-Till, Alastair~M.
  Buchan, and Iris~Q. Grunwald.
\newblock {e-ASPECTS software is non-inferior to neuroradiologists in applying
  the ASPECT score to computed tomography scans of acute ischemic stroke
  patients}.
\newblock {\em International Journal of Stroke}, 12(6):615--622, August 2017.

\bibitem{Nambiar2014}
V.~Nambiar, S.~I. Sohn, M.~A. Almekhlafi, ..., M.~Goyal, M.~D. Hill, and B.~K.
  Menon.
\newblock {CTA Collateral Status and Response to Recanalization in Patients
  with Acute Ischemic Stroke}.
\newblock {\em American Journal of Neuroradiology}, 35(5):884--890, May 2014.

\bibitem{Nielsen2018}
Anne Nielsen, Mikkel~Bo Hansen, Anna Tietze, and Kim Mouridsen.
\newblock {Prediction of Tissue Outcome and Assessment of Treatment Effect in
  Acute Ischemic Stroke Using Deep Learning}.
\newblock {\em Stroke}, 49(6):1394--1401, June 2018.

\bibitem{Nogueira2017}
Raul~G. Nogueira, Ashutosh~P. Jadhav, Diogo~C. Haussen, ..., David~S.
  Liebeskind, Jeffrey~L. Saver, and Tudor~G. Jovin.
\newblock {Thrombectomy 6 to 24 Hours after Stroke with a Mismatch between
  Deficit and Infarct}.
\newblock {\em New England Journal of Medicine}, 378(1):11--21, January 2018.

\bibitem{Olivot2009}
Jean~Marc Olivot, Michael Mlynash, Vincent~N. Thijs, ..., Roland Bammer,
  Michael~P. Marks, and Gregory~W. Albers.
\newblock {Optimal tmax threshold for predicting penumbral tissue in acute
  stroke}.
\newblock {\em Stroke}, 40(2):469--475, 2009.

\bibitem{Pexman2001}
J.~H~W Pexman, P.~A. Barber, M.~D. Hill, ..., M.~E. Hudon, W.~Y. Hu, and A.~M.
  Buchan.
\newblock {Use of the Alberta Stroke Program Early CT Score (ASPECTS) for
  assessing CT scans in patients with acute stroke}.
\newblock {\em American Journal of Neuroradiology}, 22(8):1534--1542, 2001.

\bibitem{Pinto2018a}
Adriano Pinto, S{\'{e}}rgio Pereira, Raphael Meier, ..., Roland Wiest,
  Carlos~A. Silva, and Mauricio Reyes.
\newblock {Enhancing Clinical MRI Perfusion Maps with Data-Driven Maps of
  Complementary Nature for Lesion Outcome Prediction}.
\newblock In {\em MICCAI}, pages 107--115. Spinger, 2018.

\bibitem{Powers2018}
William~J. Powers, Alejandro~A. Rabinstein, Teri Ackerson, ..., Andrew~M.
  Southerland, Deborah~V. Summers, and David~L. Tirschwell.
\newblock {2018 Guidelines for the Early Management of Patients With Acute
  Ischemic Stroke: A Guideline for Healthcare Professionals From the American
  Heart Association/American Stroke Association}.
\newblock {\em Stroke}, 49(3), March 2018.

\bibitem{Puetz2009}
V.~Puetz, I.~Dzialowski, M.~D. Hill, and A.~M. Demchuk.
\newblock {The Alberta Stroke Program Early CT Score in Clinical Practice: What
  have We Learned?}
\newblock {\em International Journal of Stroke}, 4(5):354--364, October 2009.

\bibitem{Ribo2007}
Marc Ribo, Carlos~A. Molina, Pilar Delgado, ..., Alex Rovira, Josep Munuera,
  and Jose Alvarez-Sabin.
\newblock {Hyperglycemia during Ischemia Rapidly Accelerates Brain Damage in
  Stroke Patients Treated with tPA}.
\newblock {\em Journal of Cerebral Blood Flow {\&} Metabolism},
  27(9):1616--1622, September 2007.

\bibitem{Riggs1963}
Helena~E. Riggs and Charles Rupp.
\newblock {Variation in Form of Circle of Willis}.
\newblock {\em Archives of Neurology}, 8(1):8, January 1963.

\bibitem{Robben2020}
David Robben, Anna~M.M. Boers, Henk~A. Marquering, ..., Aad van~der Lugt, Robin
  Lemmens, and Paul Suetens.
\newblock {Prediction of final infarct volume from native CT perfusion and
  treatment parameters using deep learning}.
\newblock {\em Medical Image Analysis}, 59:101589, January 2020.

\bibitem{Robben2018a}
David Robben and Paul Suetens.
\newblock {Perfusion Parameter Estimation Using Neural Networks and Data
  Augmentation}.
\newblock In {\em MICCAI-SWITCH workshop}, pages 439--446. Springer, 2019.

\bibitem{Roy1890}
C.~S. Roy and C.~S. Sherrington.
\newblock {On the Regulation of the Blood-supply of the Brain}.
\newblock {\em The Journal of Physiology}, 11(1-2):85--158, January 1890.

\bibitem{Saver2015}
Jeffrey~L. Saver, Mayank Goyal, Alain Bonafe, ..., Richard {du Mesnil de
  Rochemont}, Oliver~C. Singer, and Reza Jahan.
\newblock {Stent-Retriever Thrombectomy after Intravenous t-PA vs. t-PA Alone
  in Stroke}.
\newblock {\em New England Journal of Medicine}, 372(24):2285--2295, June 2015.

\bibitem{Scalzo2012}
Fabien Scalzo, Qing Hao, Jeffry~R. Alger, Xiao Hu, and David~S. Liebeskind.
\newblock {Regional Prediction of Tissue Fate in Acute Ischemic Stroke}.
\newblock {\em Annals of Biomedical Engineering}, 40(10):2177--2187, October
  2012.

\bibitem{Scheldeman2021}
Lauranne Scheldeman, Anke Wouters, and Robin Lemmens.
\newblock {Imaging selection for reperfusion therapy in acute ischemic stroke
  beyond the conventional time window}.
\newblock {\em Journal of Neurology}, 359:1317--1329, October 2021.

\bibitem{Shetty2006}
Sanjay~K. Shetty and Michael~H. Lev.
\newblock {CT Perfusion (CTP)}.
\newblock In R.~Gilberto Gonz{\'{a}}lez, Joshua~A. Hirsch, W.J. Koroshetz,
  Michael~H. Lev, and Pamela~W. Schaefer, editors, {\em Acute Ischemic Stroke:
  Imaging and Intervention}, pages 87--113. Springer-Verlag, Berlin/Heidelberg,
  2006.

\bibitem{Simard2007}
J~Marc Simard, Thomas~A Kent, Mingkui Chen, Kirill~V Tarasov, and Volodymyr
  Gerzanich.
\newblock {Brain oedema in focal ischaemia: molecular pathophysiology and
  theoretical implications}.
\newblock {\em The Lancet Neurology}, 6(3):258--268, March 2007.

\bibitem{Strandgaard1973}
S.~Strandgaard, J.~Olesen, E.~Skinhoj, and N.~A. Lassen.
\newblock {Autoregulation of Brain Circulation in Severe Arterial
  Hypertension}.
\newblock {\em BMJ}, 1(5852):507--510, March 1973.

\bibitem{Symon1979}
L~Symon, N~M Branston, and O~Chikovani.
\newblock {Ischemic brain edema following middle cerebral artery occlusion in
  baboons: relationship between regional cerebral water content and blood flow
  at 1 to 2 hours.}
\newblock {\em Stroke}, 10(2):184--191, March 1979.

\bibitem{Tan2009}
I.Y.L. Tan, A.M. Demchuk, J.~Hopyan, ..., S.P. Symons, A.J. Fox, and R.I. Aviv.
\newblock {CT Angiography Clot Burden Score and Collateral Score: Correlation
  with Clinical and Radiologic Outcomes in Acute Middle Cerebral Artery
  Infarct}.
\newblock {\em American Journal of Neuroradiology}, 30(3):525--531, March 2009.

\bibitem{Tawil2017}
Salwa~El Tawil and Keith~W Muir.
\newblock {Thrombolysis and thrombectomy for acute ischaemic stroke}.
\newblock {\em Clinical Medicine}, 17(2):161--165, April 2017.

\bibitem{Thomalla2020}
G{\"{o}}tz Thomalla, Florent Boutitie, Henry Ma, ..., Ona Wu, Albert~J. Yoo,
  and Ramin Zand.
\newblock {Intravenous alteplase for stroke with unknown time of onset guided
  by advanced imaging: systematic review and meta-analysis of individual
  patient data}.
\newblock {\em The Lancet}, 396(10262):1574--1584, November 2020.

\bibitem{Thomalla2011}
G{\"{o}}tz Thomalla, Bastian Cheng, Martin Ebinger, ..., Jochen~B. Fiebach,
  Jens Fiehler, and Christian Gerloff.
\newblock {DWI-FLAIR mismatch for the identification of patients with acute
  ischaemic stroke within 4{\textperiodcentered}5 h of symptom onset
  (PRE-FLAIR): a multicentre observational study}.
\newblock {\em The Lancet Neurology}, 10(11):978--986, November 2011.

\bibitem{Thomalla2018}
G{\"{o}}tz Thomalla, Claus~Z. Simonsen, Florent Boutitie, ..., Norbert
  Nighoghossian, Salvador Pedraza, and Christian Gerloff.
\newblock {MRI-Guided Thrombolysis for Stroke with Unknown Time of Onset}.
\newblock {\em New England Journal of Medicine}, 379(7):611--622, August 2018.

\bibitem{Todd1986}
N~V Todd, P~Picozzi, A~Crockard, and R~W Russell.
\newblock {Duration of ischemia influences the development and resolution of
  ischemic brain edema.}
\newblock {\em Stroke}, 17(3):466--471, May 1986.

\bibitem{VanderEecken1953}
Henry~M. {Vander Eecken} and Raymond~D. Adams.
\newblock {The Anatomy and Functional Significance of the Meningeal Arterial
  Anastomoses of the Human Brain}.
\newblock {\em Journal of Neuropathology {\&} Experimental Neurology},
  12(2):132--157, April 1953.

\bibitem{Vila2000}
Nicolas Vila, Jose Castillo, Antonio Davalos, and Angel Chamorro.
\newblock {Proinflammatory Cytokines and Early Neurological Worsening in
  Ischemic Stroke}.
\newblock {\em Stroke}, 31(10):2325--2329, October 2000.

\bibitem{VonKummer2017}
R{\"{u}}diger von Kummer and Imanuel Dzialowski.
\newblock {Imaging of cerebral ischemic edema and neuronal death}.
\newblock {\em Neuroradiology}, 59(6):545--553, June 2017.

\bibitem{Wang2000}
Yang Wang, Weixing Hu, Alejandro~D. Perez-Trepichio, ..., Anthony~J. Furlan,
  Anthony~W. Majors, and Stephen~C. Jones.
\newblock {Brain Tissue Sodium Is a Ticking Clock Telling Time After Arterial
  Occlusion in Rat Focal Cerebral Ischemia}.
\newblock {\em Stroke}, 31(6):1386--1392, June 2000.

\bibitem{Wheeler2015a}
Hayley~M. Wheeler, Michael Mlynash, Manabu Inoue, ..., Greg Zaharchuk, Matus
  Straka, and Gregory~W. Albers.
\newblock {The Growth Rate of Early DWI Lesions is Highly Variable and
  Associated with Penumbral Salvage and Clinical Outcomes following
  Endovascular Reperfusion}.
\newblock {\em International Journal of Stroke}, 10(5):723--729, July 2015.

\bibitem{White1998}
Richard~P. White, Colin Deane, Patrick Vallance, and Hugh~S. Markus.
\newblock {Nitric Oxide Synthase Inhibition in Humans Reduces Cerebral Blood
  Flow but Not the Hyperemic Response to Hypercapnia}.
\newblock {\em Stroke}, 29(2):467--472, February 1998.

\bibitem{White1999}
Richard~P. White, Claire Hindley, Peter~M. Bloomfield, ..., Patrick Vallance,
  David~J. Brooks, and Hugh~S. Markus.
\newblock {The Effect of the Nitric Oxide Synthase Inhibitor L-NMMA on Basal
  CBF and Vasoneuronal Coupling in Man: A PET Study}.
\newblock {\em Journal of Cerebral Blood Flow {\&} Metabolism}, 19(6):673--678,
  June 1999.

\bibitem{Wintermark2006}
Max Wintermark, Adam~E. Flanders, Birgitta Velthuis, ..., Julien Bogousslavsky,
  William~P. Dillon, and Salvador Pedraza.
\newblock {Perfusion-CT Assessment of Infarct Core and Penumbra}.
\newblock {\em Stroke}, 37(4):979--985, April 2006.

\bibitem{Winzeck2018}
Stefan Winzeck, Arsany Hakim, Richard McKinley, ..., Greg Zaharchuk, Roland
  Wiest, and Mauricio Reyes.
\newblock {ISLES 2016 and 2017-Benchmarking Ischemic Stroke Lesion Outcome
  Prediction Based on Multispectral MRI}.
\newblock {\em Frontiers in Neurology}, 9(SEP), September 2018.

\bibitem{Wise1983}
R.~J.~S. Wise, S.~Bernardi, R.~S.~J. Frackowiak, N.~J. Legg, and T.~Jones.
\newblock {Serial Observations on the Pathophysiology of Acute Stroke}.
\newblock {\em Brain}, 106(1):197--222, 1983.

\bibitem{Wouters2016}
Anke Wouters, Patrick Dupont, Soren Christensen, ..., Greg Albers, Vincent
  Thijs, and Robin Lemmens.
\newblock {Association Between Time From Stroke Onset and Fluid-Attenuated
  Inversion Recovery Lesion Intensity Is Modified by Status of Collateral
  Circulation}.
\newblock {\em Stroke}, 47(4):1018--1022, April 2016.

\bibitem{Wouters2021}
Anke Wouters, David Robben, Soren Christensen, ..., Gregory~W. Albers, Paul
  Suetens, and Robin Lemmens.
\newblock {Prediction of Stroke Infarct Growth Rates by Baseline Perfusion
  Imaging}.
\newblock {\em Stroke}, 53(January), September 2021.

\bibitem{Wu2006}
Ona Wu, S{\o}ren Christensen, Niels Hjort, ..., G{\"{o}}tz Thomalla, Joachim
  R{\"{o}}ther, and Leif {\O}stergaard.
\newblock {Characterizing physiological heterogeneity of infarction risk in
  acute human ischaemic stroke using MRI}.
\newblock {\em Brain}, 129(9):2384--2393, September 2006.

\bibitem{Wu2001}
Ona Wu, Walter~J. Koroshetz, Leif {\O}stergaard, ..., Lee~H. Schwamm, Robert~M.
  Weisskoff, and A.~Gregory Sorensen.
\newblock {Predicting Tissue Outcome in Acute Human Cerebral Ischemia Using
  Combined Diffusion- and Perfusion-Weighted MR Imaging}.
\newblock {\em Stroke}, 32(4):933--942, April 2001.

\bibitem{Yaghi2021}
Shadi Yaghi, Seena Dehkharghani, Eytan Raz, ..., Maarten~G. Lansberg,
  Gregory~W. Albers, and Adam de~Havenon.
\newblock {The Effect of Hyperglycemia on Infarct Growth after Reperfusion: An
  Analysis of the DEFUSE 3 trial}.
\newblock {\em Journal of Stroke and Cerebrovascular Diseases}, 30(1), January
  2021.

\bibitem{Yonas1996}
H~Yonas, R~P Pindzola, and D~W Johnson.
\newblock {Xenon/computed tomography cerebral blood flow and its use in
  clinical management.}
\newblock {\em Neurosurgery clinics of North America}, 7(4):605--16, October
  1996.

\end{thebibliography}
